\documentclass[prl,nosuperscriptaddress,twocolumn,aps,nobibnote,amssymb,floatfix,a4paper,10pt]{revtex4-2}

\usepackage{amsfonts,amsmath,amssymb,amsthm,array,booktabs,enumerate,enumitem,hyperref,pgfplots,qcircuit,mathtools,multirow,tabularx,thmtools,tikz, makecell}
\usepackage[normalem]{ulem}

\hypersetup{colorlinks=true,linkcolor=blue,citecolor=blue,urlcolor=blue}
\pgfplotsset{compat=1.18} 
\def\A{ {\cal A} }
\def\H{ {\cal H} }
\def\L{ {\cal L} }

\def\Re{\operatorname{Re}}
\def\Im{\operatorname{Im}}

\newcommand{\bra}[1]{\langle {#1} |}
\newcommand{\ket}[1]{| {#1} \rangle}
 
\newcommand{\ketbra}[2]{\ensuremath{\left|#1\right\rangle\!\!\left\langle#2\right|}}
\newcommand{\braket}[2]{\ensuremath{\!\left\langle#1|#2\right\rangle}\!}

\newcommand{\diag}[1]{\mathrm{diag}\left( #1 \right)}

\newcommand{\ot}{\otimes}
\newcommand{\tr}{\mathrm{Tr}}
\newcommand{\defeq}{\coloneqq}

\let\idotless\i 
\renewcommand{\i}{\ensuremath{\mathbf{i}}}
\renewcommand{\vec}[1]{\mathrm{vec}(#1)}

\newcommand{\lrb}[1]{\left ( #1 \right )}

\renewcommand{\i}{\ensuremath{\mathbf{i}}}

\newcommand{\mypoly}[1]{\operatorname{poly} \lrb{#1}}

\newcommand{\mylog}[1]{\operatorname{log} \lrb{#1}}
\newcommand{\mylogpow}[2]{\operatorname{log}^{#1} \lrb{#2}}

\newcommand{\myO}[1]{O\lrb{#1}}
\newcommand{\mytO}[1]{\tilde{O}\lrb{#1}}
\newcommand{\myOmega}[1]{\Omega\lrb{#1}}
\newcommand{\myTheta}[1]{\Theta\lrb{#1}}

\newenvironment{aligns}{\subequations \align} {\endalign \endsubequations}

\theoremstyle{plain}
\newtheorem{result}{Result}
\newtheorem{definition}{Problem}

\begin{document}


\title{Quantum Koopman Algorithms}

\author{David Jennings}
\affiliation{PsiQuantum, 700 Hansen Way, Palo Alto, CA 94304, USA}

\author{Kamil Korzekwa}
\affiliation{PsiQuantum, 700 Hansen Way, Palo Alto, CA 94304, USA}

\author{Matteo Lostaglio}
\email[Lead author email: ]{mlostaglio@psiquantum.com}
\affiliation{PsiQuantum, 700 Hansen Way, Palo Alto, CA 94304, USA}

\author{Guoming Wang}
\affiliation{PsiQuantum, 700 Hansen Way, Palo Alto, CA 94304, USA}

\begin{abstract} 

Solving high-dimensional dynamical systems by encoding their state in the amplitudes of a quantum state is expected to be a core application of fault-tolerant quantum computers. However, the conditions under which this paradigm yields quantum advantage can be restrictive, motivating complementary approaches. Here, we introduce Quantum Koopman Algorithms, where quantum amplitudes encode approximately closed sets of observables for a linear or nonlinear dynamics. These amplitudes evolve under the Koopman operator, and we establish conditions for efficient quantum algorithms addressing both dynamical and spectral problems. With this approach, we show that classes of open-system dynamics involving exponentially large fermionic systems can be efficiently simulated on a quantum computer. For nonlinear dynamics, we introduce a novel nonlinear interaction picture that enables quantum simulation beyond existing weak-nonlinearity conditions, and extract late-time spectral data where quantum phase estimation is unavailable.
\end{abstract}

\maketitle

\renewcommand{\thefootnote}{\arabic{footnote}}
\setcounter{footnote}{0}


\paragraph{Introduction.}

From its conception, the idea of quantum computing revolved around the possibility of efficiently representing exponentially large dynamical systems by encoding them in superposition~\cite{feynman1982simulating, lloyd1996universal}. While initially developed for the purpose of solving quantum dynamics, the same toolkit has been recently extended to classical dynamical systems~\cite{berry2014high,berry2017quantum,berry2024quantum,an2023linear,jennings2023cost,an2026quantum,PhysRevLett.133.230602,jin2022quantum, wang2026sign, an2024laplace,aftab2026linear}, which also face the curse of dimensionality in high-impact applications such as nonlinear kinetic plasma and fluid dynamics modeling~\cite{li2025potential, penuel2024detailed, jennings2025end-to-end,jennings2026simulating,ameri2023quantum,vaszary2024solving, berntson2026end, christensen2026improved}. 

The common paradigm underlying these approaches is to encode the state $x$ of a dynamical system in the amplitudes of a quantum state: $\ket{x(t)} = \sum_i x_i(t) \ket{i}$. The vector $x$ satisfies equations of motion
\begin{align}
    \label{eq:equationofmotion}
    \dot{x} = F(x), \quad x(t) \in \mathbb{C}^d, \quad t\in [0,T],
\end{align}
where $F : \mathbb{C}^d \mapsto \mathbb{C}^d$ can be a linear or nonlinear function. Quantum computers are limited to unitary evolution, hence constructing a sequence of quantum gates that transforms $\ket{x(0)}$ into $\ket{x(T)}$ typically requires a series of problem transformations and embeddings into a larger space. Whatever the strategy one follows, once $\ket{x(T)}$ is obtained, observable properties need to be extracted by quantum measurements.

This approach has been successful for some applications, but  challenges include the efficient encoding of the equations of motion on a quantum computer, and the efficient extraction of information from the quantum encoding of the solution. Here, we introduce a complementary observable-based quantum Koopman framework that offers new modalities to obtain quantum speedups for the simulation of dynamical systems. While the mathematical foundations go back to the 1930s~\cite{koopman1931hamiltonian}, Koopman-based classical methods have recently developed into practical tools in engineering and industrial applications, including aerospace, mechanics, fluid dynamics, and energy systems~\cite{budivsic2012applied,mezic2013analysis,susuki2016applied,duran2025control}. Here, these tools are repurposed for quantum computing. We amplitude-encode distinguished observables whose evolution is (approximately) closed under the dynamics, and leverage the specific strengths of quantum computing to address regimes where the evolution of observables is classically intractable. Crucially, the framework applies to nonlinear classical dynamics, but it also unifies results in quantum dynamics, e.g., for estimating dense observables~\cite{liu2024dense} or shadow Hamiltonian simulations and related algorithms~\cite{somma2025shadow, stroeks2025solving}. 

Rather than replacing state-based approaches, the observable-space formulation provides complementary routes to quantum advantage. We illustrate this by achieving \emph{exponential improvements} for classes of open quantum systems, compared to existing state-based methods. Another core challenge is that existing efficiency guarantees for quantum algorithms for nonlinear dynamics typically impose restrictive weak-nonlinearity conditions~\cite{liu2021efficient, wu2024quantum, jennings2025quantum}. We introduce a nonlinear interaction picture in which the perturbative expansion is instead organized around a solvable nonlinear flow. This yields efficient quantum algorithms in regimes where conventional weak-nonlinearity conditions can fail by an arbitrarily large margin. A final issue that we address is the limited tools available for the end-to-end analysis of dynamical systems. Here, we contribute to this toolbox by developing algorithms for extracting spectral properties of the observables' dynamics, paralleling the role of QPE and its generalizations~\cite{kitaev1995quantum,cleve1998quantum,griffiths1996semiclassical,somma2019quantum, lin2022heisenberg} in state-based approaches. We first illustrate this for the decay rates in the correlation functions of open quantum systems. Crucially, our results extend quantum spectral methods to the nonlinear regime: we show how to efficiently sample late-time oscillatory modes of nonlinear dynamics by preparing a Kaiser-windowed nonuniform history state of the Koopman evolution.


\paragraph{Formalism.} 

Given a scalar observable \(f:\mathbb C^d\to\mathbb C\), the evolution of $f(x(t))$ under Eq.~\eqref{eq:equationofmotion} is governed by the linear Koopman equation
\begin{align}
\label{eq:koopman_revised}
    \dot {f} = \tilde{K} f,
    \qquad
    \tilde{K} = \sum\nolimits_{i=1}^d F_i(x)\tfrac{\partial}{\partial x_i}
    =: F\!\cdot\!\nabla,
\end{align}
where \(\tilde{K}\) is the \emph{Koopman generator}~\footnote{Here we treat $x \in \mathbb{C}^d$ as a real vector in $\mathbb{R}^{2d}$}. Crucially, it is a linear evolution even if the original dynamics~\eqref{eq:equationofmotion} is not. The problem at hand is to extract a classically intractable observable $\mathcal{A}$.

A problem admits a QKA if it has these features:
\begin{enumerate}
  
    \item \textbf{Dynamical closure.} A large set of observables $g=\{g_m(x(t))\}_{m=1}^M$ that allows one to estimate the observable~$\A$, and whose dynamics admit an exact or approximate finite-dimensional closure:
    \begin{align}
        \dot{g} = K g + b,
    \label{eq:koopmanproblem}
    \end{align} 
    with $b\in \mathbb{C}^M$ and $K\in \mathbb{C}^{M\times M}$ obtained from a truncation of $\tilde{K}$ to the span of $\{g_m\}$.
  
    \item \textbf{Efficient encodings.} An efficient unitary state-preparation \(U_0\) for the quantum state
    \begin{equation}
        \ket{\psi(0)} \propto \sum_{m= 1}^M g_m (x(0)) \ket{m}.
    \label{eq:koopmandynamic}
    \end{equation}
    Also, an efficient unitary block-encoding \(U_{ K}\) of~\(K\)~\footnote{For a matrix $A$, this is an efficient quantum circuit $U_A$ encoding $A/\alpha$ in its top-left block, with $\alpha >0$ being a scale factor~\cite{lin2022lecture}.}, and an efficient state preparation unitary~\(U_b\) of $\ket{b}$.
    \item  \textbf{Efficient decoding.} The observable $\mathcal{A}$ can be efficiently extracted from the quantum output $\ket{\psi(t)}$ or a history state~\footnote{Here a history-state encodes dynamical data over a time interval. It takes the form $\sum_k  a_k\ket{\psi(t_k)}\ket{t_k}$, where the second register encodes a discrete set of times $t_k$ over the interval, the first register encodes the solution at that time $t_k$, and $a_k$ are associated norm-weightings due to potentially non-unitary dynamics.}.
\end{enumerate}

The resulting quantum-encoded Koopman system can then be analyzed from the point of view of its dynamic or spectral properties:

\smallskip

\textbf{Dynamic-QKA} -- a quantum algorithm outputting a coherent observable-space encoding of the initial value problem~\eqref{eq:koopmanproblem} for $\ket{\psi(0)}$ as in~\eqref{eq:koopmandynamic}, and extracting the observable $\A$ from the coherent output.

\textbf{Spectral-QKA} -- a quantum algorithm determining a spectral property $\A$ of $K$. $K$ encodes, e.g., information about invariance and ergodicity, coherent oscillations, growth or decay rates~\cite{budivsic2012applied}. 

\smallskip


\paragraph{QKAs for linear dynamics.} 

Consider exponentially large linear ODEs with $F(x) = Ax$, and take an exponential set of linear observables, $g_m(x) = w_m^\dag x$, or quadratic ones, $g_m(x) = x^\dag O_m x$. Dynamical closure requires $A^\dag w_m = \sum_{m'} K^*_{mm'} w_{m'}$ for the former, and $A^\dag O_m +  O_m A = \sum_{m'} K_{mm'} O_{m'}$ for the latter, where $K_{mm'}$ are complex coefficients. Assuming the exact closure, we get $\dot{g} = K g$. Dynamic-QKAs solve this initial value problem by encoding~$g$ into a quantum state~\eqref{eq:koopmandynamic} and evolving it via quantum ODE solvers~\cite{berry2014high,berry2017quantum,berry2024quantum,an2023linear,jennings2023cost,an2026quantum}. This setting includes Hamiltonian simulation~\cite{lloyd1996universal,berry2007efficient,berry2015simulating}, dense output simulation~\cite{liu2024dense} and shadow Hamiltonian simulation~\cite{somma2025shadow} as special cases, see Appendix~\ref{appendix:prior}. Spectral-QKAs, by contrast, reconstruct the eigenvalues of~$K$ via quantum phase estimation (QPE) and generalizations~\cite{kitaev1995quantum,cleve1998quantum,gilyen2019quantum,low2019hamiltonian}. 


\paragraph{Open dynamics of exponentially many free fermions.}

We demonstrate the above in the context of fermionic open quantum systems (a similar approach applies to bosonic systems). Free fermionic systems model materials and electronic transport~\cite{groth2014kwant,kloss2021tkwant}. General quantum algorithms for many-body systems require $\mathrm{poly}(N)$ resources in the number $N$ of modes (or particles), which is matched by a classical algorithm~\cite{valiant2001quantum,terhal2002classical}. At the same time, numerical simulations for large $N$ become prohibitive, especially for 3D systems, raising the question of whether more efficient quantum algorithms exist. 

Refs.~\cite{somma2025shadow, stroeks2025solving} introduced quantum algorithms for closed free fermionic systems with an exponentially improved $\mathrm{polylog}(N)$ complexity. Here, we deal with the open quantum system case, yielding exponential improvements over prior methods. Introduce Majorana operators $c=(c_1, \dots, c_{2N})$, the quadratic Hamiltonian $H= \frac{\i}{4} c^\dag h c$, with $h^T = -h$, and jump operators  $L_\mu = l_\mu\cdot c$, with~$l_\mu$ denoting row vectors of length $2N$. A natural choice of observables are quadratic Majorana operators $O_{ij}=\frac{\i}{2}[c_i,c_j]$, whose expectation values are the elements of the covariance matrix $\Gamma$, evolving as (see Appendix~\ref{app:fermions}): 
\begin{align}
    \label{eq:matrix-ODE}
     \dot{\Gamma} = B \Gamma + \Gamma B^\top + Y,\qquad B=h-X, \quad
\end{align}
with $X=2\sum_\mu\Re(l_\mu^\dagger l_\mu)$, $ Y=-4\sum_\mu\Im(l_\mu^\dagger l_\mu)$. Define:

\begin{definition}[Structured free fermion problem]
\label{problem:freefermions}
Take Eq.~\eqref{eq:matrix-ODE} with $h$ and $\{l_{\mu i}\}$ satisfying:
 \begin{enumerate}
     \item \emph{Locality:} sparsity $O(\mathrm{polylog}(N))$, with the positions of nonzero elements and their values efficiently specifiable and of size $\Theta(1)$.
     \item \emph{Extensive or point couplings:} $\Theta(N)$ or $\Theta(1)$ nonzero elements of $h$, $\Re(l_{\mu i})$ and $\Im(l_{\mu i})$.
 \end{enumerate}
\end{definition}

These assumptions are expected to hold in ordinary local many-body systems, where each mode participates in $\Theta(1)$ couplings in a geometrically structured way, with extensive or boundary dissipation, and with intensive couplings that do not grow with the system size. Examples include tight-binding chains~\cite{alba2023logarithmic}, Kitaev wires~\cite{van2019dynamical} and quadratic fermions on a lattice with bulk local dissipation~\cite{prosen2008third}, or boundary-driven transport models~\cite{zunkovic2010explicit}.

We present several algorithms for Problem~\ref{problem:freefermions}, see Appendix~\ref{appendix:proofs_fermions} for details. First, we show that exponentially many free fermions admit efficient quantum simulation:

\begin{result}[Exponentially large covariance matrix] 
\label{res:covariance}
    For Problem~\ref{problem:freefermions}, set a target time $t= O(\mathrm{polylog}(N))$, a target error $\epsilon = O(1/\mathrm{polylog}(N))$, and assume the Frobenius norm $\|\Gamma(\tau)\|_F=\Theta(\sqrt{N})$ for $\tau \in [0,t]$~\footnote{Physically, the condition on $\|\Gamma\|_F$ is generic, as it holds whenever a finite fraction of fermionic modes remains a constant away from the infinite temperature state throughout the evolution.}. Then, under the assumption that one can efficiently prepare $\ket{\psi(0)}$, there is a quantum algorithm involving $O(\mathrm{polylog}(N))$ elementary gates that outputs a quantum state $\epsilon$-close to 
    \begin{align}
    \label{eq:finalstate}
        \ket{\psi(t)} \propto \sum_{i,j} \Gamma_{ij}(t) \ket{ij}. 
    \end{align}
\end{result}

Crucially, the quantum encoded covariance matrix obtained in Result~\ref{res:covariance} supports efficient extraction of physically relevant thermodynamic and spectral quantities. First, we construct a Dynamic-QKA outputting an $\epsilon$-additive estimate of the total dissipated heat per fermion, using $O(\mathrm{polylog}(N))$ elementary gates. Moreover, we can estimate decay rates (a spectral property~$\A$) in the case where the open dynamics satisfies time-translation symmetry, i.e., $[h, X]=0$. This is the well-known secular approximation and it is common in microscopic derivations of Markovian master equations~\cite{breuer2002theory, fleming2010rotating, lostaglio2017markovian}. Under this assumption, $ \Gamma(t)$ has decay rates $\nu_{kj}:= \nu_k + \nu_j$ (including zero), where $\{\nu_{j}\}$ are the eigenvalues of $X$, with corresponding eigenstates $\{\ket{\phi_j}\}$. The weight of each decay channel is controlled by the overlap $a_{kj}(0) = \bra{\phi_k} \otimes \bra{\phi_j} \ket{\mathrm{vec}(\Gamma(0))}$. We then construct an importance-sampling Spectral-QKA giving an $\epsilon$-additive approximation of the decay rate $\nu_{kj} $ for the fermionic covariance matrix dynamics with probability equal to the spectral weight $|a_{kj}(0)|^2$, with each run requiring $\mathrm{polylog}(N)$ elementary quantum gates. Finally, in this setting, we also show how to efficiently prepare the steady state covariance matrix, encoded as a unitary or a quantum state, at a cost scaling with the inverse of the spectral gap of the Lindbladian and polylogarithmic in~$N$.


\paragraph{QKAs \& the nonlinear interaction picture.} 

Most practically important classical dynamical systems, including those in fluid and plasma dynamics, are governed by nonlinear equations of motion. Since quantum computers implement linear evolution, the problem must be first embedded into a larger linear system of observables. Controlling the dynamical closure error resulting from such a finite-dimensional embedding is a central bottleneck preventing the quantum simulation of the highly relevant strongly nonlinear regime~\cite{forets2017explicit,chen2024carleman, liu2021efficient,jennings2025quantum,jennings2025end-to-end,bravyi2025quantum, novikau2025quantum}.

One prominent closure scheme is the Carleman embedding method, which tracks the monomial observables $\{g_m\} \equiv \{x^{\otimes k}\}_{k=1}^{N_C}$ up to some truncation order $N_C$. For quadratic nonlinear dynamics~\footnote{Note that any polynomial nonlinearity can be reduced to a quadratic one.}, 
\begin{equation}
    \dot{x} = F_1 x + F_2 x^{\otimes 2},
\end{equation}
it has been shown that if 
\begin{equation}
    R_C := \tfrac{\|F_2\| \|x(0)\|}{-\mu(F_1)} < 1,
    \end{equation}
where $\mu(F_1)<0$ is the log-norm of $F_1$, then the dynamical closure error decreases exponentially in $N_C$, and an efficient quantum algorithm for monomial observables can be constructed~\cite{liu2021efficient, costa2023further}. This, and other structurally similar conditions~\cite{jennings2025quantum}, all require the strength of the nonlinearity to be small, severely limiting the class of nonlinear dynamics admitting efficient quantum algorithms~\cite{Lewis2024limitationsquantum, novikau2025globalizing, joseph2023quantum}.

We tackle this key constraint through the lens of the interaction picture in quantum algorithms, where one splits a Hamiltonian into one piece that is dominant, but can be fast-forwarded, and another `perturbative' one. In the Koopman framework, an analogous separation yields
\begin{equation}
    \label{eq:interaction_p}
    \dot{x} = F(x) = A(x) + \delta B(x), \quad x(t) \in \mathbb{C}^d,
\end{equation}
where $A(x)$ generates `tractable' nonlinear dynamics and $\delta B(x)$ is a perturbation around this flow. Specifically, we assume that a set of observables $\eta:=\{\eta_j(x)\}$ exists whose span is invariant under the dynamics of the Koopman generator $\tilde{K}_A:= A(x) \cdot \nabla$. Examples include settings where the Koopman modes of~$\tilde{K}_A$ are known, such as for the logistic equation or the Burgers' equation~\cite{page2018koopman}.

We then construct a truncated hierarchy of equations for $g:=\{\eta^{\otimes k}(x)\}_{k=1}^{N_C}$, and analyze the dynamical closure error $\epsilon_K$ via the $R$--number above, or its generalizations.  A key component of this construction is that Koopman eigenmodes of $\tilde{K}_A$ generate a lattice \footnote{More precisely, if $\eta_1(x(t))$ and $\eta_2(x(t))$ are eigenmodes of $\tilde{K}_A$ with eigenvalues $\lambda_1$ and $\lambda_2$ respectively, then $\eta_1(x(t)) \eta_2(x(t))$ is an eigenmode of $\tilde{K}_A$ with eigenvalue $\lambda_1+\lambda_2$.}.  In the traditional interaction picture, $\delta B$ may generate weak off-diagonal terms in the eigenbasis of $A$. Our scheme can be viewed as a nonlinear interaction picture, where $A$ gives rise to a \emph{block-}diagonal generator $\tilde{K}_A$, with the perturbation $\delta B$ corresponding to weak higher-order block contributions. Thus, the truncation error is controlled by the nonlinearity of the perturbation $\delta B(x)$, rather than by the full nonlinearity of $F$: nonlinearity is measured relative to the solvable reference flow $A(x)$. We illustrate this formalism with a simple example in Appendix~\ref{app:nonlinear_ip}. With this scheme, we can construct amplitude encodings of $g$ in quantum states and provide a QKA for observables $\A$ of interest, for classes of strongly nonlinear systems.


\paragraph{Applications of the nonlinear interaction picture.} 

To illustrate this method (see Appendix~\ref{app:karleman_k}), consider a nonlinear population model given by Eq.~\eqref{eq:interaction_p} with 
\begin{equation}
\label{Eq:populationdynamics}
    A(x)_i = r_ix_i(1-\tfrac{x_i}{X_i}),\; \; \delta B(x)_i = -x_i^2\sum_{j,k} \! J_{i,jk} \eta_j \eta_k,
\end{equation}
where $\eta_j:=(X_j-x_j)/x_j$. The unperturbed term $A(x)$ is a logistic growth for population, with growth rate $r_i$ and carrying capacity $X_i$; the perturbation $\delta B(x)$ introduces mutualistic or competitive quadratic interactions through the unused-resource variables $\eta_j(x)$. This model can exhibit chaotic behavior. In the non-interacting case $J_{i,jk}\equiv 0$, the variables $\eta_i$ are Koopman eigenmodes of the nonlinear flow. We construct the following nonlinear interaction picture quantum algorithm:

\begin{result}[Exponentially large nonlinear population dynamics]
    \label{res:populations}
   Assume that $r_i>0$, $X_i>0$. Given a circuit that prepares the  initial state $\ket{\eta(0)}$, standard sparse access assumptions on $J_{i,jk}$, and the condition
    \begin{align}
    \label{eq:RK}
    R_{\mathrm{K}} = \tfrac{\|G_2\|\|\eta(0)\|}{\min_i r_i} < 1,
    \end{align}
    where $[G_2]_{i,jk}= X_i J_{i,jk}$,
    there is a quantum algorithm involving $O(\mathrm{polylog}(d))$ gates that outputs a quantum state that encodes in superposition the entire trajectory $\eta(t)$ for a time $t= \mathrm{polylog}(d)$.
\end{result}

Compared with conventional Carleman quantum algorithms, our construction offers several advantages. First, for Carleman, no a priori analytical guarantee like \eqref{eq:RK} is available for this class of systems. Second, our numerical investigations show that the regime of applicability of Carleman is much more limited than our approach. Fig.~\ref{fig:convergence} illustrates this behavior for a three-population system. Finally, in Appendix~\ref{app:karleman_separation_r_numbers}, we construct classes of dynamics with provably unbounded $R$-number separation, i.e., we analytically prove that there exist families of dynamical systems for which the conventional Carleman $R$-number diverges while the nonlinear-interaction-picture $R$-number vanishes~\footnote{With the usual caveat that the $R$-number is a sufficient but not necessary criterion of convergence}.

\begin{figure}
    \centering
    \includegraphics[width=0.49\linewidth]{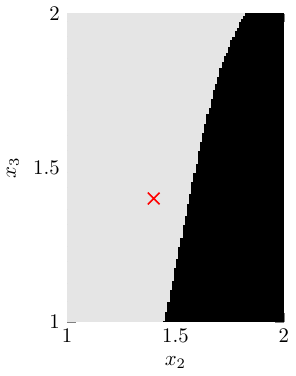}
    \includegraphics[width=0.49\linewidth]{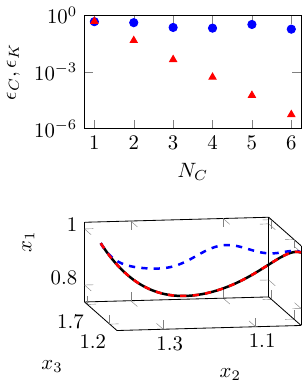}
    \caption{\textbf{Carleman and nonlinear interaction picture convergence.} Left: Region of Carleman convergence (gray) and lack thereof (black) for a 3D system described by Eq.~\eqref{Eq:populationdynamics} with parameters $(r_i,X_i,J_{i,jk})$ as in Appendix~\ref{app:karleman_k} and $x(0)=[1,x_2,x_3]$. The nonlinear interaction picture numerically converges (error at $N_C=3$ is smaller than $N_C=1$) everywhere in the presented region.  Top right: Dependence on truncation errors $\epsilon_C$ for Carleman (blue circles) and $\epsilon_K$ for the nonlinear interaction picture (red triangles) on $N_C$ for $x(0)=[1,1.4,1.4]$ (red cross on the left). Truncation error is the maximal Euclidean distance between the exact and approximate state during the whole evolution for $t\in[0,0.1]$ (enough for equilibration). Bottom right: Exact (black), Carleman  (blue dashed), and nonlinear interaction picture  (red dashed) trajectories for $x(0)$, $N_C=6$, and $t\in[0,0.1]$. 
    }
    \label{fig:convergence}
\end{figure}


\paragraph{Spectral-QKA \& nonlinear late-time dynamics.} 

Nonlinear dynamics can display late-time persistent oscillations, and extracting the frequencies of such oscillatory modes is a non-trivial task. Here, we introduce a spectral-QKA to perform importance sampling of late-time oscillatory modes for a class of nonlinear systems. More precisely, take a finite-dimensional normal Koopman matrix $K$ with eigenvalues $\lambda_k$ and eigenvectors $\ket{\phi_k}$. Assume stability, $\mathrm{Re}(\lambda_k) \leq 0$ for all $k$. Given initial-time Koopman data $\ket{g(0)}$, we expand it into `oscillatory' and `decaying' modes
\begin{align}
\label{eq:initialdataspectral}
    \!\!\! \ket{g(0)} =\!\! 
    \underbrace{\sum_{k| \mathrm{Re}(\lambda_k) =0} a_k(0) \ket{\phi_k}}_{\mathrm{oscillatory \; modes}}
    +
    \!\!\underbrace{\sum_{k| \mathrm{Re}(\lambda_k) <0} a_k(0) \ket{\phi_k}}_{\mathrm{decaying \; modes}}.
\end{align}
Although QPE cannot be applied directly since $K$ is not (anti)Hermitian, after the decaying modes are suppressed we can prepare the nonuniform history state
\[
    \sum_{j=0}^{J-1}
    \beta_j\ket{j}\ket{g(T_1+j\Delta t)},
\]
where $\beta_j$ are Kaiser-window coefficients. A linear-system embedding prepares this state efficiently, while the Kaiser window exponentially suppresses Fourier tails from off-grid frequencies~\cite{berry2024analyzing, greenaway2024case, patel2026optimal}. Applying an inverse QFT then yields importance sampling of the oscillatory frequencies. More generally, the same construction prepares history states with arbitrary efficiently preparable weights. We obtain (proofs in Appendix~\ref{app:spectralqka}):
\begin{result}[Importance sampling of oscillatory modes]
\label{res:spectralinformal}
Assume access to a block-encoding of a normal Koopman matrix $K$ with scale factor $\alpha$, and a unitary procedure to prepare $\ket{g(0)}$ as in \eqref{eq:initialdataspectral}, with an $\Omega(1)$ fraction of oscillatory modes. Then, there exists an importance-sampling quantum algorithm whose output is $O(\zeta)$-close in total variation distance to the ideal outcome obtained by sampling an oscillatory mode $k$ according to its spectral weight and estimating $\Im(\lambda_k)$ to additive error $\epsilon$ with probability at least $1-\delta$. The query complexity of the algorithm is
\begin{align}
\mytO{
    \alpha
    \left(
        \frac{1}{\Delta}\log\frac{1}{\zeta}
        +
        \frac{1}{\epsilon}\log\frac{1}{\delta}
    \right)
    \mylogpow{2}{\frac{1}{\zeta}}
},
\end{align}
where
$\Delta$ is the spectral gap
\begin{align}
    \Delta
    =
    -\max_{k|\mathrm{Re}(\lambda_k)<0}
    \mathrm{Re}(\lambda_k).
\end{align}
\end{result}


\paragraph{Outlook.} 

We introduced Quantum Koopman Algorithms (QKAs), a framework for simulating linear and nonlinear dynamics on a quantum computer by solving the initial-value problem for a set of generalized observables (Dynamic-QKA), or performing modal analysis of the Koopman operator (Spectral-QKA). Since the seminal Applied Koopmanism work~\cite{budivsic2012applied}, Koopman-based techniques have evolved from a promising conceptual direction into practical tools for modeling, control, forecasting, and diagnosis, especially in settings with large data~\cite{zhang2025data, williams2015data}. In contrast, the extension to the quantum regime is a novel frontier, with exponential computational space but constrained access to raw data. Together, these results establish observable-space dynamics as a distinct route to quantum advantage, extending efficient quantum simulation and spectral analysis to regimes inaccessible to existing state-based approaches.

\paragraph{Author contributions.}  
  
Authors are listed alphabetically. DJ developed the initial theory of the nonlinear interaction picture, proposed using spectral methods for nonlinear dynamics, and contributed to the theory of Result~\ref{res:covariance}. KK was responsible for Result~\ref{res:covariance} (Lindblad dynamics of free fermions and the constructions of related efficient block-encodings and state preparations) and Result~\ref{res:populations} (nonlinear interaction picture results and the population dynamics example). ML contributed to setting out the framework, Results~\ref{res:covariance} and \ref{res:populations},  was responsible for Results~\ref{res:heat}, \ref{res:decay}, \ref{res:steady} in the Supplemental Material, and developed an earlier version of Result~\ref{res:spectralinformal}. GW established Result~\ref{res:R_separation} in the Supplemental Material (unbounded $R$-number separation) and developed Result~\ref{res:spectralinformal} (spectral QKA) in its current form. All authors wrote the paper.

\paragraph{Acknowledgments.}
We thank all our colleagues at PsiQuantum for useful conversations and feedback, and in particular Alice Barthe and Burak Şahinoğlu. ML thanks Gabriele Benedetti for useful discussions on Koopman theory.
After completion of this work, Ref.~\cite{simon2026efficientquantumalgorithmlinear} appeared, which is related to our results on fermionic open system dynamics.

\nocite{krovi2023improved,shukla2024efficient,dalzell2024shortcut,somma2025quantum,an2024fast,casas2016asymptotically,yang2026towards,grover2005fixedpoint,yoder2014fixed,martyn2021grand}

\bibliographystyle{apsrev4-1} 
\bibliography{references}

\clearpage
\onecolumngrid

\setcounter{section}{0}
\setcounter{figure}{0}
\renewcommand{\thesection}{S\arabic{section}}
\renewcommand{\theequation}{S\arabic{equation}}
\renewcommand{\thefigure}{S\arabic{figure}}
\renewcommand{\thetable}{S\arabic{table}}

\renewcommand{\theHequation}{S\arabic{equation}}
\renewcommand{\theHfigure}{S\arabic{figure}}
\renewcommand{\theHtable}{S\arabic{table}}
\renewcommand{\theHsection}{S\arabic{section}}

\begin{center}
\textbf{\large Supplemental Material}
\end{center}
\setcounter{equation}{0}


\section{Relation to prior works}
\label{appendix:prior}

\begin{figure}[h]
    \centering
    \includegraphics[width=0.5\linewidth]{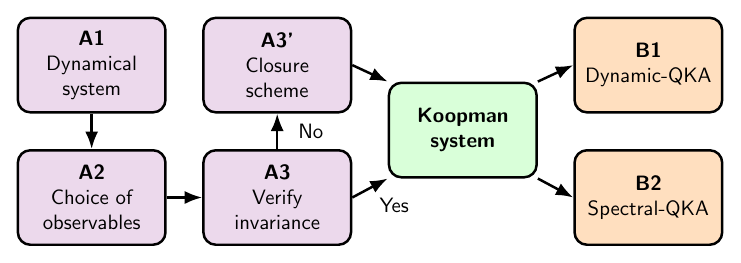}
    \caption{\textbf{How to construct a QKA.} 
    }
    \label{fig:Koopmanscheme}
\end{figure}

A schematic construction of a QKA is shown in Fig.~\ref{fig:Koopmanscheme}. One first specifies a dynamical system, i.e., a vector field~$F$ in Eq.~\eqref{eq:equationofmotion} (A1), and then selects a family of observables $\{g_m\}$ (A2), which induces the Koopman evolution from Eq.~\eqref{eq:koopman_revised}. If these observables span an invariant subspace of $\tilde{K}$, one obtains a finite-dimensional system directly~(A3). Otherwise, one introduces an appropriate closure scheme to obtain the effective dynamics from Eq.~\eqref{eq:koopmanproblem} (A3$'$).

\emph{Hamiltonian simulation~\cite{lloyd1996universal,berry2007efficient,berry2015simulating}:} 
In A2, we can specialize to the following linear observables:
\begin{align}
\label{eq:canonicalobservables}
   g_m(x) = e_m^\dag x, \quad n\in\{1, \dots, M=d\},
\end{align}
where $e_m$ are canonical basis states. Then, $ \dot{g}_m  = (-i H g)_m$, which can be compactly written simply as the Schr\"odinger equation $\dot{x} = -i H x$. The invariance condition is trivially satisfied and we reduce to standard Hamiltonian simulation, with a Koopman operator $K=-iH$.

\emph{Dense output simulation~\cite{liu2024dense}:} We keep A1 to be the Schr\"odinger dynamics, but for A2  we introduce the following observables:
 \begin{aligns}
     f_{ij}(x) &= x^\dag \ketbra{j}{i} x, \\\ 
     J(x)  &= \int_0^t x^\dag(\tau) O x(\tau) d\tau,
     \label{eq:denseobservables}
\end{aligns}
for some matrix $O$ of interest. 
The $M=d^2+1$ observables at hand can be compactly described as $g(x) = [x^* \otimes x,J(x)]^T$. Then, we can readily work out the Koopman operator $K$ satisfying $\dot{g} = K g$:
  \begin{align}
      K = \begin{bmatrix}
          \i (H^* \otimes I - I \otimes H) & 0 \\
          \textrm{vec}(O) & 0
      \end{bmatrix},
      \label{eq:koopmandense}
  \end{align}
  where $\textrm{vec}(O)$ is the vectorized version of $O$.
The invariance property is trivially satisfied and we reduce to the dense output algorithm of Ref.~\cite{liu2024dense}. Solving the resulting ODE system is efficient under appropriate conditions (see, e.g., Ref.~\cite{jennings2023cost}), which here pose conditions on $|J(t)|$ and the Hilbert-Schmidt norm of $O$.

\emph{Shadow Hamiltonian simulation~\cite{somma2025shadow}:} For the special case where $A = -\i H$, with $H$ Hermitian, and with quadratic observables, the invariance condition reads $- \i [H,O_m] = \sum_m K_{mm'} O_{m'}$, which is the commutator closure of Ref.~\cite{somma2025shadow}. When $K=-\i h$ with $h$ Hermitian, then $h$ is the shadow Hamiltonian. 


\section{Lindblad dynamics of free fermionic systems}
\label{app:fermions}

The evolution of the density operator $\rho$ under general Lindblad dynamics is given by the following linear equation of motion:
\begin{equation}
    \dot{\rho} = \H[\rho]+\L[\rho],
\end{equation}
where
\begin{aligns}
    \H[\rho] := & -\i[H,\rho],\\
    \L[\rho] := & \sum_\mu \left(L_\mu \rho L_\mu^\dagger - \frac{1}{2}\{L_\mu^\dagger L_\mu,\rho\}\right).    
\end{aligns}
In the above, $H$ is the Hamiltonian describing closed quantum evolution, whereas $L_\mu$ are jump operators that couple the system to its environment. The evolution of a generic observable $A_i$ is governed by the adjoint superoperators:
\begin{equation}
    \dot{A}_i  = \H^\dagger[A_i] +\L^\dagger[A_i].
\end{equation}

We consider an open dynamics of a system of $N$ free fermions with linear couplings to the bath~\cite{prosen2008third}, where $N$ can be exponential, $N=2^n$.  Its Hamiltonian $H$ is quadratic in Majorana operators $c_i$,
\begin{equation}
    H = \frac{\i}{4}c^\dagger h  c = \frac{\i}{4} \sum_{i,j=1}^{2N} h_{ij} c_i c_j,
\end{equation}
and satisfies the antisymmetric property $h_{ij}=-h_{ji}$, whereas the jump operators $L_\mu$ are given by:
\begin{equation}
    L_\mu = l_\mu\cdot c = \sum_{i=1}^{2N} l_{\mu,i} c_i.
\end{equation}
Let us recall that $c_i$ are Hermitian and satisfy the canonical anticommutation relation $\{c_i,c_j\}=2\delta_{ij} I$, from which one can derive the following useful commutation relations:
\begin{aligns}
    [c_i,c_j]&=2(1-\delta_{ij}) c_i c_j, \\
    [c_ic_j,c_k] &= 2(\delta_{jk} c_i -  \delta_{ik} c_j),\\
    [c_i,c_jc_k] &= 2(\delta_{ij} c_k -  \delta_{ik} c_j),\\
    [c_ic_j,c_kc_l] &= 2(\delta_{jk}c_ic_l -\delta_{jl}c_ic_k+\delta_{ik}c_lc_j-\delta_{il}c_kc_j).
\end{aligns}
The set of $k$-th degree fermionic observables is the linear span of all operators that are products of exactly $k$ Majorana operators, $c_{i_1}\dots c_{i_k}$, where all indices are distinct. 

We first focus on a set of quadratic observables
\begin{equation}
    A_{kl}:=\left\{ \begin{matrix}
        \i c_k c_l && \text{for $k\neq l$,}\\      
        0 && \text{for $k= l$.}
    \end{matrix}
    \right.
\end{equation}
Note that the imaginary constant $\i$ was used in this definition so that the expectation values $\langle A_{kl}\rangle$ yield entries of the covariance matrix $\Gamma$:
\begin{equation}
\label{eq:gamma}
    \Gamma_{kl} := \frac{\i}{2} \langle [c_k,c_l] \rangle = \i \langle c_k c_l\rangle (1-\delta_{kl}) = \langle A_{kl} \rangle,
\end{equation}
which fully specifies the state when it is a Gaussian state. The evolution of $A_{kl}$ reads
\begin{equation}
    \dot{A}_{kl} = \H^\dagger[A_{kl}] +\L^\dagger[A_{kl}],
\end{equation}
with the Hamiltonian part given by:
\begin{align*}
    \H^\dagger[A_{kl}] &= \frac{\i}{4} \sum_{i, j} h_{ij}[c_k c_l,c_i c_j]\\
    &= \frac{\i}{2}\sum_{i}  (h_{il} c_i c_k - h_{ik} c_i c_l + h_{li} c_k c_i -h_{ki} c_l c_i) \\
    &= \frac{\i}{2}\sum_{i}  (h_{il} [c_i, c_k] - h_{ik} [c_i, c_l])\\
    &= \i\sum_{i}  (h_{il} (1-\delta_{ik})c_ic_k - h_{ik} (1-\delta_{il})c_ic_l)\\
    &= \i\sum_{i,j} (h_{il} (1-\delta_{ik})\delta_{jk} - h_{ik} (1-\delta_{il})\delta_{jl})c_ic_j\\
    &= \i \sum_{i,j} (h_{il} \delta_{jk} - h_{ik} \delta_{jl}) (1-\delta_{ij}) c_i c_j\\
    &= \sum_{i,j} (h_{il} \delta_{jk} - h_{ik} \delta_{jl}) A_{ij},
\end{align*}
and the dissipative parts described by:
\begin{align*}
    \L^\dagger_\mu[A_{kl}] &= \i\sum_{ij}l_{\mu,i}^* l_{\mu,j} c_ic_kc_lc_j  -\frac{\i}{2}\sum_{ij}l_{\mu,i}^* l_{\mu,j} (c_ic_jc_kc_l +c_kc_lc_ic_j) \\
    &= \frac{\i}{2}\sum_{ij}l_{\mu,i}^* l_{\mu,j} (c_i[c_k c_l,c_j] + [c_i,c_k c_l]c_j) \\
    &=\i\sum_{ij}l_{\mu,i}^* l_{\mu,j} \left(c_i(\delta_{lj} c_k -  \delta_{kj} c_l) + (\delta_{ik} c_l -  \delta_{il} c_k)c_j\right) \\
    &= \i\sum_{i}\left( l_{\mu,i}^* l_{\mu,l} c_i c_k   - l_{\mu,i}^* l_{\mu,k} c_i c_l + l_{\mu,k}^* l_{\mu,i} c_l c_i   - l_{\mu,l}^* l_{\mu,i}   c_k c_i \right)\\
    &= 2\i\sum_{i\neq k} \Re(l_{\mu,i}^* l_{\mu,l}) c_i c_k -2\i\sum_{i\neq l} \Re(l_{\mu,i}^* l_{\mu,k}) c_i c_l   - 4\Im(l_{\mu,k}^* l_{\mu,l})I\\
    &= 2\i\sum_{i} \left( \Re(l_{\mu,i}^* l_{\mu,l}) c_i c_k - \Re(l_{\mu,i}^* l_{\mu,k}) c_i c_l\right)   - 4\Im(l_{\mu,k}^* l_{\mu,l})I\\
    &= 2\i\sum_{i,j} \left( \Re(l_{\mu,i}^* l_{\mu,l})\delta_{jk}  - \Re(l_{\mu,i}^* l_{\mu,k})\delta_{jl}\right)c_i c_j   - 4\Im(l_{\mu,k}^* l_{\mu,l})I\\
    &= 2\sum_{i,j} \left( \Re(l_{\mu,i}^* l_{\mu,l})\delta_{jk}  - \Re(l_{\mu,i}^* l_{\mu,k})\delta_{jl}\right)A_{ij}   - 4\Im(l_{\mu,k}^* l_{\mu,l})I.
\end{align*}
We thus conclude that
\begin{equation}
    \dot{A}_{kl} = \sum_{i,j}\left((h_{il} \delta_{jk} - h_{ik} \delta_{jl})+2\sum_\mu \left( \Re(l_{\mu,i}^* l_{\mu,l})\delta_{jk}  - \Re(l_{\mu,i}^* l_{\mu,k})\delta_{jl}\right)\right)A_{ij}- 4\sum_\mu \Im(l_{\mu,k}^* l_{\mu,l})I.
\end{equation}

The set of the above evolution equations for all observables $A_{kl}$ can now be concisely written as a matrix equation. To that end, we first collect all observables into a $2N \times 2N$ operator-valued matrix $A$:
\begin{equation}
    A = \sum_{k,l=1}^{2N} \ketbra{k}{l} \ot A_{kl} = \i\begin{pmatrix}
        0 &c_1c_2&c_1c_3&\dots&c_1c_{2N}\\
        c_2c_1 & 0 & c_2c_3 & \dots & c_2 c_{2N}\\
        \vdots & \vdots & \vdots & \ddots & \vdots\\
        c_{2N}c_1 & c_{2N}c_2 & c_{2N} c_3 &\dots & 0
    \end{pmatrix}.
\end{equation}
Then, we also introduce $2N\times 2N$ matrices $X_\mu$ and $Y_\mu$:
\begin{equation}
     X_\mu  = 2\Re(l_\mu^\dagger l_\mu),\qquad
     Y_\mu  = -4\Im(l_\mu^\dagger l_\mu),
\end{equation}
whose matrix elements are the coefficients appearing in the dissipative part of the evolution:
\begin{equation}
    X_{\mu,ij}  = 2\Re(l_{\mu,i}^* l_{\mu,j}),\qquad
    Y_{\mu,ij}  = -4\Im(l_{\mu,i}^* l_{\mu,j}),
\end{equation}
Finally, we also introduce
\begin{equation}
        X = \sum_\mu X_\mu,\qquad Y := \sum_\mu Y_\mu,
\end{equation}
as well as
\begin{equation}
    B = h - X,
\end{equation}
and now can write the evolution concisely as
\begin{align}
    \dot{A} &= (B\ot I) A +A(B^\top\ot I)  +Y\ot I.
\end{align}
Taking the expectation value of the above, we arrive at the equation of motion for the covariance matrix:
\begin{equation}
    \dot{\Gamma} = B\Gamma + \Gamma B^\top + Y.
\end{equation}

Instead of dealing with a matrix equation for $A$, we can also vectorize it, creating $(2N)^2$-dimensional operator-valued vector $\vec{A}$:
\begin{equation}
    \vec{A} = \sum_{k,l=1}^{2N} \ket{kl}\otimes A_{kl} = \begin{pmatrix}
        0,&c_1c_2,&c_1c_3,&\dots,&c_{2N}c_{2N-1},&0
    \end{pmatrix}^\top.
\end{equation}
Then, the evolution of $\vec{A}$ can be inferred from that of $A$ by using the standard vectorization identity,
\begin{equation}
    \vec{CAD}=C\ot D^\top \vec{A},
\end{equation} 
as well as the antisymmetry $h^\top=-h$ and symmetry $X^\top=X$. Using these, we arrive at
\begin{equation}
    \frac{d\vec{A}}{dt} = (\mathbb{B}\ot I) \vec{A} +\vec{Y}\ot I,
\end{equation}
where
\begin{equation}
    \mathbb{B} = 
        B \ot I +I\ot B,
\end{equation}
and $\vec{Y}$ is just a vectorization of $Y$:
\begin{equation}
    \vec{Y} = \sum_{ij} Y_{ij} \ket{ij}.
\end{equation}
We can now again take the expectation value to arrive at the equation of motion for the vectorized covariance matrix
\begin{equation}
    \frac{d \vec{\Gamma}}{dt} = \mathbb{B} \vec{\Gamma} +\vec{Y},
\end{equation}
and so we obtain just a driven linear equation for the elements of the covariance matrix.

We then see that the set of quadratic Majorana observables extended with an identity observable $I$ is invariant under open dynamics described by a quadratic Hamiltonian and linear jump operators. Note that the extension by $I$ is unnecessary when coefficient of jump operators, $l_{\mu,i}$, are all real. In fact, in such a case quadratic observables are simply invariant due to $Y=0$. Moreover, for all $K$, the set of observables of even degree smaller or equal to $2K$ is also invariant under such dynamics. To see this, consider the $2K$-th order fermionic observable
\begin{equation}
    A_{i} =  c_{i_1}\dots c_{i_{2K}},\qquad i:= [i_1,\dots,i_{2K}],
\end{equation}
with all indices distinct and note that its evolution can be written as
\begin{align}
    \dot{A}_i & =  \frac{d(c_{i_1}c_{i_2})}{dt} c_{i_3}\dots c_{i_{2K}} + \dots + c_{i_1}\dots c_{i_{2K-2}}\frac{d(c_{i_{2K-1}}c_{i_{2K}})}{dt}.
\end{align}
We see that each of $K$ terms contains $(2K-2)$ Majorana operators and the derivative of a pair of Majorana operators. As we have shown above, such a derivative is given by a sum of pairs of Majoranas and an identity operator. In other words, $\dot{A}_{i}$ can be written as a sum of $2K$ or $2K-2$ Majorana operators. Then, it can happen that some Majoranas will have the same index and will thus square to identity, thus decreasing the degree by 2. Hence, the final expression for $\dot{A}_i$ will be given by a sum of terms consisting of $2k$ Majorana operators for $k\in\{0,\dots,K\}$.


\section{Technical details on results for the structured free fermion problem}
\label{appendix:proofs_fermions}


\subsection{Proof of Result~1}


\subsubsection{Query complexity cost}

Using the results in Ref.~\cite{krovi2023improved, jennings2023cost}, the complexity $Q$ for outputting a state $\epsilon$-close to Eq.~\eqref{eq:finalstate} is 
\begin{align}
    Q=O\left( \alpha  \max_{\tau \in [0,t]} \|e^{\mathbb{B} \tau}\| \ \mathfrak{g}(t) t \ \log \frac{ \max\{t, \| \mathrm{vec}(Y)\|\}}{\epsilon \Gamma_{\min}(t)} \ \log \frac{\mathfrak{g}(t) t}{\epsilon}\right),
\end{align}
where $\alpha$ is the block-encoding prefactor for $\mathbb{B}$ and
\begin{subequations}
\begin{align}
    \mathfrak{g}(t)  & := \max_{\tau \in [0,t]} \frac{\| \Gamma(\tau) \|_F}{\|\Gamma(t)\|_F},\\
    \Gamma_{\min}(t) &:= \min_{\tau \in [0,t]} \|\Gamma(\tau)\|_F.
\end{align}
\end{subequations}
The complexity $Q$ measures the number of times one needs to apply the unitary block-encoding of $\mathbb{B}$ and a unitary preparing a quantum state proportional to $\textrm{vec}(Y)$. In what follows, we first bound all terms, except for $\alpha$ and $\|\vec{Y}\|$, with $\mathrm{polylog}(N)$. Then, we show that $\alpha = \mathrm{polylog}(N)$,  $\|\vec{Y}\|=O(\sqrt{N})$, and that both unitaries can be constructed with $\mathrm{polylog}(N)$ gates under our assumptions. 

The norm $\|\mathbb{B}\|$ sets the timescale for the problem. This is readily computed, and found to obey 
\begin{equation}
  \|\mathbb{B}\| = O(\| h\|+\| X\|) = O(\mathrm{polylog}(N)),
\end{equation}
where we used Gershoring's circle theorem, the fact that the sparsity of $h$ and $X$ is $\mathrm{polylog}(N)$, and the largest elements are $\Theta(1)$ in absolute value. Next, we bound the quantity $\|e^{\mathbb{B} t}\|$. Let $U$ be the unitary diagonalizing $X$, $X= U \Lambda U^{\dag}$, and $\tilde{h} := U^\dag h U$. Then 
\begin{align}
    \|e^{\mathbb{B} t}\| = \| e^{B t}\|^2 = \| e^{(\tilde{h} - \Lambda)t}\|^2 \leq e^{2\mu(\tilde{h} - \Lambda)t},
\end{align}
where $\mu(A)$ is the log-norm of $A$, $\mu(A) = \lambda_{\max}\left[\frac{A+A^\dag}{2}\right]$. Since
\begin{align}
    \mu(\tilde{h} - \Lambda) = \mu(-\Lambda) = -\lambda_{\min}(X) \leq 0, 
\end{align}
and $X\geq 0$, we conclude that 
\begin{align}
     \|e^{\mathbb{B} t}\| \leq 1.
\end{align}
Also, by assumption 
\begin{align}
    \mathfrak{g}(t) = \Theta(1),
\end{align}
and $\Gamma_{\min}(t) = \Theta(\sqrt{N})$. Hence, we end up with
\begin{align}
      Q = \alpha  \times \log \max\{t, \| \mathrm{vec}(Y)\| \} \times O(\mathrm{polylog}(N)).
\end{align}


\subsubsection{Efficient block-encoding of \texorpdfstring{$\mathbb{B}$}{B}}
\label{app:fermions_block}

In order to efficiently block-encode $\mathbb{B}$, it is sufficient to block-encode $h$ and $X$, which can then be combined using an LCU. It is known that a sparse $h$ can be efficiently block-encoded under the assumptions of Result~1, so we only need to block-encode $X$:
\begin{equation}
    X = 2\sum_\mu \Re(l_\mu^\dagger l_\mu).
\end{equation}
To achieve this, we will construct a block-encoding $U_{X'}$ of
\begin{equation}
    X'=\sum_\mu l_\mu^\dagger l_\mu,
\end{equation}
because from it and its complex conjugate, using an LCU, we can obtain the block-encoding of $X$. We assume that a matrix $l_{\mu i}$ is $s$-sparse, so that we can introduce the following index functions,
\begin{aligns}
    I_\mu(s_1):&\quad \text{index of $s_1$-th nonzero entry of $l_{\mu,(\cdot)}$},\label{eq:index1}\\
    J_i(s_2):&\quad \text{index of $s_2$-th nonzero entry of $l_{(\cdot),i}$},\label{eq:index2}
\end{aligns}
where $s_1,s_2\in\{1,\dots s\}$. We now assume access to the following standard oracles for positions of nonzero elements:
\begin{aligns}
    O_I\ket{\mu,s_2} & = \ket{\mu,I_\mu(s_2)},\\
    O_J\ket{s_1,i} & = \ket{J_i(s_1),i}.
\end{aligns}
We also need access to the values of $l_{\mu i}$. With the access assumed in Result~1, we can efficiently construct the following standard oracles (we need to assume proper scaling so that $|l_{\mu i}|\leq 1)$:
\begin{subequations}
\begin{align}
    V\ket{0,\mu,i} &= \left(l_{\mu i}\ket{0}+\sqrt{1-|l_{\mu i}|^2}\ket{1}\right) \ot \ket{\mu,i},\\
    V^*\ket{0,\mu,i} &= \left(l_{\mu i}^*\ket{0}+\sqrt{1-|l_{\mu i}|^2}\ket{1}\right) \ot \ket{\mu,i}.
\end{align}
\end{subequations}
Finally, we will also make use of a uniform state preparation $H_n$ over $n$ states~\cite{shukla2024efficient}:
\begin{equation}
    H_n\ket{0} =  \frac{1}{\sqrt{n}} \sum_{n_1=0}^{n-1} \ket{n_1}.
\end{equation}

The following circuit $U_{X'}$ block-encodes $X'$ with a block-encoding prefactor $1/s^2$:
\begin{equation}
   \begin{tikzpicture}[baseline=(img.center)]
        \node[anchor=south west, inner sep=0] (img) at (0,0)
            {\includegraphics[height=0.275\textwidth]{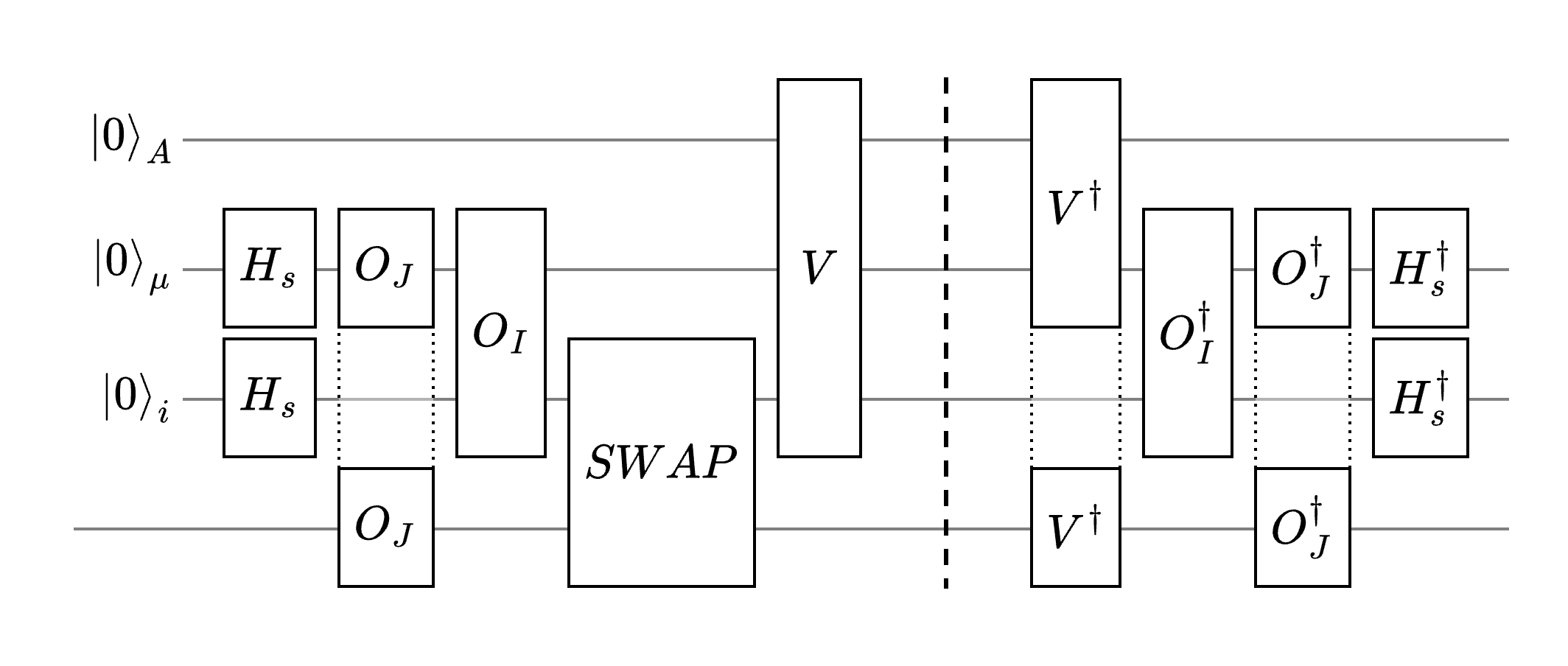}};

        \begin{scope}[x={(img.south east)},y={(img.north west)}]

            \node at (1,0.788) {$\bra{0}_A$};
            \node at (1,0.593) {$\bra{0}_\mu$};
            \node at (1,0.395) {$\bra{0}_i$};
        \end{scope}
    \end{tikzpicture}
\end{equation}
To see this, let us follow the evolution of kets from the initial state to the dashed line, then the evolution of bras from the final state to the dashed line, and finally we will take the inner product. Evolution of kets:
\begin{align*}
    \ket{0,0,0,k} &\xrightarrow{H_s^{\ot 2}} \frac{1}{s}\sum_{s_1,s_2=0}^{s-1} \ket{0,s_1,s_2,k} \xrightarrow{O_J} \frac{1}{s}\sum_{s_1,s_2=0}^{s-1} \ket{0,J_k(s_1),s_2,k} \xrightarrow{O_I} \frac{1}{s}\sum_{s_1,s_2=0}^{s-1} \ket{0,J_k(s_1),I_{J_k(s_1)}(s_2),k}\\
    &\xrightarrow{SWAP} \frac{1}{s}\sum_{s_1,s_2=0}^{s-1} \ket{0,J_k(s_1),k,I_{J_k(s_1)}(s_2)}\xrightarrow{V} \frac{1}{s}\sum_{s_1,s_2=0}^{s-1} l_{J_k(s_1),k}\ket{0,J_k(s_1),k,I_{J_k(s_1)}(s_2)} + \ket{1}\ot\text{junk}.
\end{align*}
Evolution of bras:
\begin{align*}
    \bra{0,0,0,l} &\xrightarrow{(H_s^\dagger)^{\ot 2}} \frac{1}{s}\sum_{s_1',s_2'=0}^{s-1} \bra{0,s_1',s_2',l} \xrightarrow{O_J^\dagger} \frac{1}{s}\sum_{s_1',s_2'=0}^{s-1} \bra{0,J_l(s_1'),s_2',l} \xrightarrow{O_I^\dagger} \frac{1}{s}\sum_{s_1',s_2'=0}^{s-1} \bra{0,J_l(s_1'),I_{J_l(s_1')}(s_2'),l}\\
    &\xrightarrow{V^\dagger} \frac{1}{s}\sum_{s_1',s_2'=0}^{s-1} l^*_{J_l(s_1'),l}\bra{0,J_l(s_1'),I_{J_l(s_1')}(s_2'),l} + \bra{2}\ot\text{junk},
\end{align*}
where,  with a slight abuse of notation, we denoted by $V^\dagger$ above is the Hermitian conjugate of $V$, but when the junk output is flagged by state $\ket{2}$, not~$\ket{1}$.  Inner product:
\begin{align}
   \bra{0,0,0,l}U_{X'}\ket{0,0,0,k} & = \frac{1}{s^2}\sum_{s_1,s_1',s_2,s_2'=0}^{s-1}(l^*_{J_l(s_1'),l}\bra{0,J_l(s_1'),I_{J_l(s_1')}(s_2'),l})(l_{J_k(s_1),k}\ket{0,J_k(s_1),k,I_{J_k(s_1)}(s_2)})\\
   & = \frac{1}{s^2}\sum_{s_1,s_1',s_2,s_2'=0}^{s-1} l_{J_l(s_1'),l}^*l_{J_k(s_1),k} \bra{J_l(s_1')}J_k(s_1)\rangle \bra{I_{J_l(s_1')}(s_2')}k\rangle \bra{l}I_{J_k(s_1)}(s_2)\rangle\\
   & = \frac{1}{s^2}\sum_{s_1,s_1',s_2,s_2'=0}^{s-1} l_{J_l(s_1'),l}^*l_{J_k(s_1),k} \delta_{J_l(s_1'),J_k(s_1)} \bra{I_{J_k(s_1)}(s_2')}k\rangle \bra{l}I_{J_l(s_1')}(s_2)\rangle\\
   & = \frac{1}{s^2}\sum_{s_1,s_1'=0}^{s-1} l_{J_l(s_1'),l}^*l_{J_k(s_1),k} \delta_{J_l(s_1'),J_k(s_1)}. 
\end{align}
The above is exactly $X_{lk}/s^2$, as can be seen from the following:
\begin{equation}
    X_{lk}' = \sum_{\mu=1}^N l_{\mu,l}^* l_{\mu,k} = \sum_{\mu=1}^N \left(\sum_{s_1'=0}^{s-1} \delta_{\mu,J_l(s_1')}l_{\mu,l}^* \right) \left(\sum_{s_1=0}^{s-1} \delta_{\mu,J_k(s_1)}l_{\mu,k}\right) = \sum_{s_1,s_1'=0}^{s-1} l_{J_l(s_1'),l}^* l_{J_k(s_1),k} \delta_{J_l(s_1'),J_k(s_1)}.
\end{equation}
Therefore, we conclude that one can efficiently block-encode $\mathbb{B}$.


\subsubsection{Efficient state preparation of \texorpdfstring{$\vec{Y}$}{vec(Y)}}
\label{app:fermions_blockY}

Our aim now is to construct an efficient state preparation unitary $U_{\ket{\vec{Y}}}$ of the normalized version $\ket{\vec{Y}}$ of the vector
\begin{equation}
    \vec{Y} = -4\sum_\mu \Im(l_\mu^* \ot l_\mu).
\end{equation}
If there are only $\Theta(1)$ nonzero elements of $\mathrm{Im} \, l_{\mu i}$, then $\ket{\mathrm{vec}(Y)}$ can trivially be prepared efficiently, and the norm $\| \mathrm{vec}(Y)\|$ will be $\Theta(1)$. Otherwise, when there are $\Theta(N)$ nonzero elements, we will construct a state preparation for the vector
\begin{equation}
    \vec{Y'} = \sum_\mu l_\mu^* \ot l_\mu,
\end{equation}
because from it and its complex conjugate using an LCU we can obtain the state preparation of $\ket{\vec{Y}}$.

We start by noting that although $\vec{Y'}$ has dimension $(2N)^2$, it has at most $2s^2N$ nonzero elements, due to the assumption that $l_{\mu i}$ is $s$-sparse. Using the index functions introduced in Eqs.~\eqref{eq:index1}-\eqref{eq:index2}, we can write
\begin{align}
    \vec{Y'} &= \sum_{i,j,\mu=1}^{2N} l_{\mu i}^* l_{\mu j}\ket{ij}=\sum_{i,j=1}^{2N}\ \sum_{s_1=0}^{s-1} l_{J_i(s_1),i}^* l_{J_i(s_1),j}\ket{ij}\\
    &=\sum_{i=1}^{2N}\ket{i}\ot\sum_{s_1,s_2=0}^{s-1} l_{J_i(s_1),i}^* l_{J_i(s_1),I_{J_i(s_1)}(s_2)}\ket{I_{J_i(s_1)}(s_2)}\\
    & =: \sum_{i=1}^{2N} \ket{i} \ot \|\psi_i\| \ket{\psi_i}.
\end{align}
This means that the normalized version of $\vec{Y'}$ reads
\begin{equation}
        \ket{\vec{Y'}}=\frac{1}{\|\vec{Y'}\|} \sum_{i=1}^{2N} \ket{i}\ot \|\psi_i\| \ket{\psi_i},\qquad {\|\vec{Y'}\|}^2 = \sum_{i=1}^{2N} \|\psi_i\|^2.
\end{equation}
Note that under the assumption that the coefficients of the jump operators are $\Theta(1)$, we get that $\|\vec{Y'}\|$ is $\Theta(\sqrt{N})$. Hence, $\|\vec{Y}\| = O(\sqrt{N})$. 

The following circuit then prepares a state proportional to $\ket{\vec{Y'}}$ using the oracles introduced in the previous section: 
\begin{equation}
   \begin{tikzpicture}[baseline=(img.center)]
        \node[anchor=south west, inner sep=0] (img) at (0,0)
            {\includegraphics[height=0.275\textwidth]{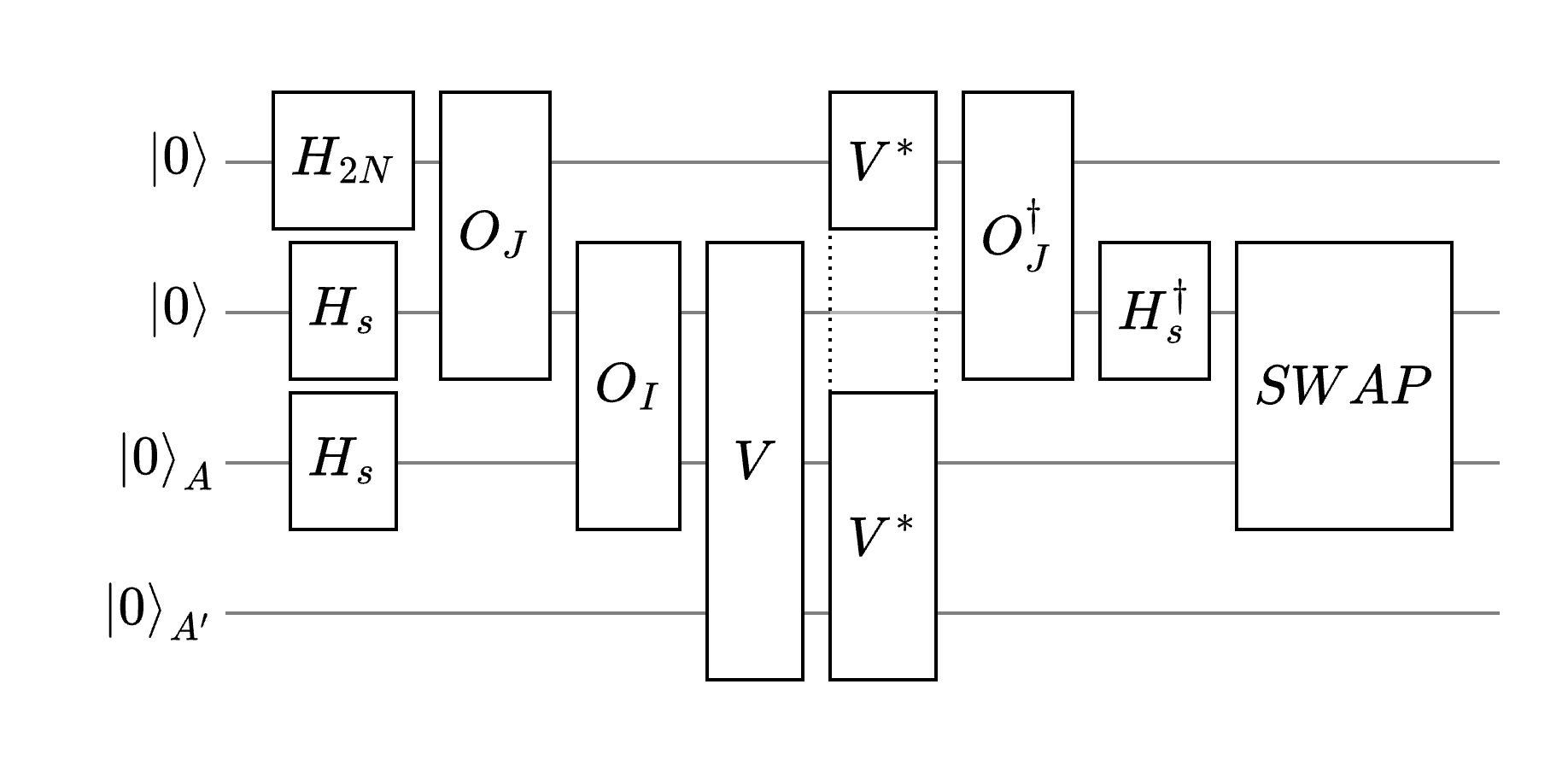}};
        \begin{scope}[x={(img.south east)},y={(img.north west)}]
            \node at (1,0.4) {$\bra{0}_A$};
            \node at (1,0.205) {$\bra{0}_{A'}$};
        \end{scope}
    \end{tikzpicture}
\end{equation}
Let us see how it is done by analyzing the above circuit gate by gate:
\begin{align}
    \ket{0,0,0,0}&\xrightarrow{H_{2N}\ot H_s^{\ot 2}} \frac{1}{s\sqrt{2N}}\sum_{i=1}^{2N}\sum_{s_1,s_2=0}^{s-1} \ket{i,s_1,s_2,0} \\
    &\xrightarrow{O_J} \frac{1}{s\sqrt{2N}}\sum_{i=1}^{2N} \sum_{s_1,s_2=0}^{s-1} \ket{i,J_i(s_1),s_2,0}\\
    & \xrightarrow{O_I} \frac{1}{s\sqrt{2N}}\sum_{i=1}^{2N} \sum_{s_1,s_2=0}^{s-1} \ket{i,J_i(s_1),I_{J_i(s_1)}(s_2),0}\\
    & \xrightarrow{V} \frac{1}{s\sqrt{2N}}\sum_{i=1}^{2N} \sum_{s_1,s_2=0}^{s-1} l_{J_i(s_1),I_{J_i(s_1)}(s_2)} \ket{i, J_i(s_1),I_{J_i(s_1)}(s_2),0} + \ket{\perp}\\
    & \xrightarrow{V^*} \frac{1}{s\sqrt{2N}}\sum_{i=1}^{2N} \sum_{s_1,s_2=0}^{s-1} l_{J_i(s_1),i}^* l_{J_i(s_1),I_{J_i(s_1)}(s_2)} \ket{i, J_i(s_1),I_{J_i(s_1)}(s_2),0} + \ket{\perp}\\
    & \xrightarrow{O_J^\dagger} \frac{1}{s\sqrt{2N}}\sum_{i=1}^{2N} \sum_{s_1,s_2=0}^{s-1} l_{J_i(s_1),i}^* l_{J_i(s_1),I_{J_i(s_1)}(s_2)} \ket{i,s_1,I_{J_i(s_1)}(s_2),0} + \ket{\perp}\\
    & \xrightarrow{H_s^\dagger} \frac{1}{s^{3/2}\sqrt{2N}}\sum_{i=1}^{2N} \sum_{s_1,s_2=0}^{s-1} l_{J_i(s_1),i}^* l_{J_i(s_1),I_{J_i(s_1)}(s_2)} \ket{i,0,I_{J_i(s_1)}(s_2),0} + \ket{\perp}\\
    & \xrightarrow{SWAP}  \frac{1}{s^{3/2}\sqrt{2N}}\sum_{i=1}^{2N} \ket{i}\ot \sum_{s_1,s_2=0}^{s-1} l_{J_i(s_1),i}^* l_{J_i(s_1),I_{J_i(s_1)}(s_2)} \ket{I_{J_i(s_1)}(s_2)} \ot \ket{00} + \ket{\perp}\\
     & = \frac{1}{s^{3/2}\sqrt{2N}}\sum_{i=1}^{2N} \ket{i}\ot \|\psi_i\| \ket{\psi_i} \ot \ket{00} + \ket{\perp}\\
     & = \frac{\|\vec{Y'}\|}{s^{3/2}\sqrt{2N}} \ket{\vec{Y'}}\ot \ket{00} + \ket{\perp}.
\end{align}
The norm of the state with ancillas in zero is
\begin{equation}
    \frac{\|\vec{Y'}\|}{s^{3/2}\sqrt{2N}}
\end{equation}
which is $\Theta(1)$  under the assumptions of Result~1. We thus conclude that we can efficiently prepare $\ket{\vec{Y}}$.


\subsection{Extracting the total dissipated heat}
\label{appendix:theorem2}

\setcounter{result}{3}
\begin{result}[Total heat dissipated] 
\label{res:heat}
 Under the assumptions of Result~1, assume $\epsilon$-multiplicative estimates of $\| h\|_F$ and $\| \Gamma(0)\|_F$ with $\epsilon=1/\mathrm{polylog}(N)$. Then, there is a quantum algorithm outputting an $\epsilon$-additive estimate of the total dissipated heat per fermion, using $O(\mathrm{polylog}(N))$ elementary gates.
\end{result}

The energy $E(\tau)$ of the system at time $\tau$ reads:
\begin{align}
    E(\tau) = \langle H \rangle  = \frac{\i}{4} \sum_{i\neq j} h_{ij} \langle c_i c_j \rangle \stackrel{\eqref{eq:gamma}}{=} \frac{1}{4} \sum_{i \neq j} h_{ij} \Gamma_{ij}(\tau) = -\frac{1}{4} \mathrm{Tr}[h \Gamma(\tau)],
\end{align}
because
\begin{equation}
    \tr(h\Gamma)=\sum_{i,j}\bra{i}h\ketbra{j}{j} \Gamma \ket{i} = \sum_{i,j}h_{ij} \Gamma_{ji} =-\sum_{i,j}h_{ij} \Gamma_{ij}. 
\end{equation}
Since $h$ is time-independent, the total dissipated heat $Q$ per fermion is just the average change of energy in the system: 
\begin{align}
    \nonumber
       - Q &= \frac{E(t) - E(0)}{N} = \frac{1}{4N} \mathrm{Tr}[h \Gamma(0)] - \frac{1}{4N} \mathrm{Tr}[h \Gamma(t)] = \frac{1}{4N} \mathrm{vec}(h)^\dag \mathrm{vec}(\Gamma(t)) - \frac{1}{4N} \mathrm{vec}(h)^\dag \mathrm{vec}(\Gamma(0)) \\
         \nonumber
        & = \frac{1}{4N} \| \mathrm{vec}(h)\| \| \mathrm{vec}(\Gamma(t))\| \braket{\mathrm{vec}(h)}{ \mathrm{vec}(\Gamma(t))} - \frac{1}{4N} \| \mathrm{vec}(h)\| \| \mathrm{vec}(\Gamma(0))\| \braket{\mathrm{vec}(h)}{ \mathrm{vec}(\Gamma(0))}
        \\
        & =   \frac{1}{4} \frac{\| h\|_F }{\sqrt{N}} \frac{\|\Gamma(t)\|_F}{\sqrt{N}} \braket{\mathrm{vec}(h)}{ \mathrm{vec}(\Gamma(t))} - \frac{1}{4} \frac{\| h\|_F}{\sqrt{N}} \frac{\|\Gamma(0)\|_F}{\sqrt{N}} \braket{\mathrm{vec}(h)}{ \mathrm{vec}(\Gamma(0))},
\end{align}
where in the last equation of the first line we used $h^\dag = - h$. Note that the case with point couplings is trivial, since we can just output $0$ as the heat dissipated per fermion. For now, we shall hence consider the case of extensive couplings.

We thus see that $Q$ is the difference of two terms, where each term is a product of three factors. In what follows, we will explain that each of these factors has magnitude $\Theta(1)$ and show how to estimate them up to an additive error $\epsilon=1/\mathrm{polylog}(N)$ using $\mathrm{polylog}(N)$ gates. As an elementary calculation shows, this way we can obtain an $\epsilon$-additive estimate of $Q$ -- with $\epsilon$ inverse polylogarithmic in $N$ -- using $\mathrm{polylog}(N)$ gates, which will prove Result~\ref{res:heat}.
    
Result~1 provides an algorithm that prepares the quantum state $\ket{\mathrm{vec}(\Gamma(t))}$. As an intermediate step, it runs a quantum linear solver algorithm on a linear system of the form 
    \begin{align}
        L y = b,
    \end{align} 
    obtained by time-discretization of 
    \begin{equation}
\label{eq:odefreefermions}
    \vec{\dot\Gamma} = \mathbb{B}\ \vec{\Gamma} + \vec{Y},\qquad
    \mathbb{B} =  B \otimes I + I \otimes B.
\end{equation}
This outputs a `history state' of the form~\cite{jennings2023cost}:
\begin{align}
    \label{eq:historystate}
    \frac{1}{\mathcal{N}} \left(\sum_{\tau=0}^{t/\Delta t-1}  \| \mathrm{vec}(\Gamma(\tau\Delta t))\| \ket{\mathrm{vec}(\Gamma(\tau \Delta t))} \ket{\tau} + \sqrt{t/\Delta t} \| \mathrm{vec}(\Gamma(t))\|  \ket{\mathrm{vec}(\Gamma(t))} \frac{1}{\sqrt{t/\Delta t}} \sum_{\tau=t/\Delta t}^{2t/\Delta t-1} \ket{\tau}\right),
\end{align}
where $\Delta t$ is a (known) time-stepping and
\begin{align}
        \mathcal{N} = \sqrt{\sum_{\tau=0}^{t/\Delta t-1} \| \mathrm{vec}(\Gamma(\tau\Delta t))\|^2 + t/\Delta t \| \mathrm{vec}(\Gamma(t))\|^2}.
\end{align}
    Note that the amplitude of the component at the final state has been increased via a standard `idling' trick incorporated in the construction of the linear system~\cite{jennings2023cost}. Also, note that $\|\vec{A}\|=\|A\|_F$ by definition of vectorization and Frobenius norm. 
 
In our setting, the condition number of $L$ scales as $\kappa = O(\mathrm{polylog}(N))$, as it implicitly follows from the proof of Result~1. There are various techniques to estimate $\mathcal{N}$ up to a $\epsilon$-multiplicative factor approximation at a cost $\mathrm{poly}(\kappa)/\epsilon$ depending on the technique~\cite{dalzell2024shortcut}. Hence, we can estimate $\mathcal{N}$ up to $\epsilon = O( 1/\mathrm{polylog}(N))$ with $\mathrm{polylog}(N)$ queries to a block-encoding of $L$ and to a unitary preparation of a quantum state proportional to $b$. The block-encodings can be constructed efficiently from those presented in the proof of Result~1 as shown, e.g., in Ref.~\cite{jennings2023cost} for a time discretization based on the truncated-Taylor series; similarly, the state preparation of a quantum state proportional to $b$ can be achieved via the preparation of a state proportional to $\Gamma(0)$, which is efficient by assumption. This means that the cost of this estimate is $\mathrm{polylog}(N)$ gates. 

Furthermore, we can get an $\epsilon$-multiplicative error estimate for $\sqrt{t/\Delta t} \| \mathrm{vec}(\Gamma(t))\|/\mathcal{N}$ at a cost $O(1/\epsilon)$ in terms of number of preparations of the quantum state $\ket{\mathrm{vec}(\Gamma(t))}$, via amplitude estimation. For an estimate with $\epsilon = O( 1/\mathrm{polylog}(N))$, this means again $\mathrm{polylog}(N)$ gates. We have $\sqrt{t/\Delta t} = O(\mathrm{polylog}(N))$ (since $t= \Theta(\mathrm{polylog}(N)$ and $\Delta t = O(\| \mathbb{B}\|^{-1}) = O(1/\mathrm{polylog}(N))$), hence this means we also have a $\epsilon$-multiplicative estimate of $\| \mathrm{vec}(\Gamma(t))\|/\mathcal{N}$. 

Putting the two estimates together, we have an $\epsilon$-multiplicative estimate of
$\mathcal{N} \times \| \mathrm{vec}(\Gamma(t))\|/\mathcal{N} = \| \mathrm{vec}(\Gamma(t))\|$ for $\epsilon = O( 1/\mathrm{polylog}(N))$ using $\mathrm{polylog}(N)$ gates. Let $\tilde{a}$ be the estimate for $\| \mathrm{vec}(\Gamma(t))\|$. What we proved is that
\begin{align}
    | \tilde{a} - \| \mathrm{vec}(\Gamma(t))\| | \leq \epsilon \| \mathrm{vec}(\Gamma(t))\|.
\end{align}
This implies we get an $\epsilon$-additive estimate for ${\| \mathrm{vec}(\Gamma(t))\| }/{\sqrt{N}}$:
\begin{align}
    \left| \frac{\tilde{a}}{\sqrt{N}} - \frac{\| \mathrm{vec}(\Gamma(t))\| }{\sqrt{N}} \right| \leq \epsilon \frac{\| \mathrm{vec}(\Gamma(t))\|}{\sqrt{N}} = \Theta(\epsilon),
\end{align}
since ${\| \mathrm{vec}(\Gamma(t))\|}/{\sqrt{N}} = \Theta(1)$ by assumption. 
By assumption, we also know $\| h \|_F$ and $\| \Gamma(0)\|_F$ up to $\epsilon$-multiplicative error. By the same reasoning as above, this implies $\epsilon$-additive estimates on ${\| h \|_F}/{\sqrt{N}}$ and ${\| \Gamma(0)\|_F}/{\sqrt{N}}$, with $\epsilon = O(1/\mathrm{polylog}(N))$. 

What is left to be estimated is the third factor, given by the overlaps $\braket{\mathrm{vec}(h)}{\mathrm{vec}(\Gamma(\tau))}$ for $\tau\in \{0,t\}$. First, let us show how to prepare $\ket{\mathrm{vec}(h)}$. By assumption, we can efficiently construct a unitary
\begin{align}
    O_h \ket{i,j}\ket{0}= \ket{i,j}\ket{\tilde{h}_{ij}},
\end{align}
where $\tilde{h}_{ij} = h_{ij}/ h_{\mathrm{max}}$, where $ h_{\mathrm{max}}= \max_{i,j}|h_{ij}|$ (the encoding is to a suitably high bit approximation). From this, we can efficiently construct the amplitude encoding
\begin{align}
    U_h \ket{i,j}\ket{0}= \tilde{h}_{ij} \ket{i,j}\ket{0} + \ket{\mathrm{junk}},
\end{align}
where $(I \otimes \bra{0})\ket{\mathrm{junk}} =0$ (e.g., see Sec.~6.1~\cite{lin2022lecture}). We also know that we can efficiently realize the unitary
\begin{align}
    O_I \ket{i,s_2} = \ket{i, I_i(s_2)},
\end{align}
where $I_i(s_2)$ is the $s_2$-th nonzero entry of $h_{i,(\cdot)}$. 
Now, consider first the case where the number of nonzero elements of $h$ is $\Theta(N)$. We can then prepare a uniform  superposition (for which efficient circuits exist~\cite{shukla2024efficient}) and apply the $O_I$ unitary followed by the $U_h$ unitary:
\begin{align}
   \frac{1}{\sqrt{2sN}} \sum_{i=0}^{2N-1} \sum_{s_2=0}^{s-1} \ket{i,s_2}   & \xrightarrow{O_I} \frac{1}{\sqrt{2sN}} \sum_{i=0}^{2N-1} \sum_{s_2=0}^{s-1} \ket{i,I_i(s_2)}   \xrightarrow{U_h}  \frac{1}{\sqrt{2sN}} \sum_{i=0}^{2N-1} \sum_{s_2=0}^{s-1} \tilde{h}_{i I_i(s_2)}  \ket{i,I_i(s_2)} \ket{0}  +  \ket{\mathrm{junk}} \\ & = \frac{\|h\|_F}{\sqrt{2sN} h_{\mathrm{max}}} \frac{1}{\|h\|_F}\sum_{i=0}^{2N-1} \sum_{s_2=0}^{s-1}  h_{i I_i(s_2)}  \ket{i,I_i(s_2)} \ket{0} +  \ket{\mathrm{junk}} \\ & =  \frac{\|h\|_F}{\sqrt{2sN} h_{\mathrm{max}}} \ket{\mathrm{vec}(h)} \ket{0} +  \ket{\mathrm{junk}}.
\end{align}
From our assumptions $\| h\|_F = \Theta(\sqrt{N})$, $h_{\mathrm{max}}= \Theta(1)$, $s=\Theta(1)$, hence with $\Theta(1)$ rounds of amplitude amplification we can prepare $\ket{\mathrm{vec}(h)}$. The case when the number of nonzero elements of $h$ is $\Theta(1)$ is analogous, just instead of preparing a uniform superposition over $Ns$ states, one prepares a uniform superposition over a constant number of nonzero elements. This shows we can prepare $\ket{\mathrm{vec}(h)}$ with $\mathrm{polylog}(N)$ gates. We showed in Result~1 how to prepare $\ket{\mathrm{vec}(\Gamma(t))}$ with $\mathrm{polylog}(N)$ gates.  Hence a Hadamard test allows us to estimate the (real-valued) overlap $\braket{\mathrm{vec}(h)}{ \mathrm{vec}(\Gamma(t))}$ up to an additive error $\epsilon$ that is inverse polylogarithmic in $N$. The same holds for  $\braket{\mathrm{vec}(h)}{ \mathrm{vec}(\Gamma(0))}$.

Given that all terms involved can be estimated up to $\epsilon$-additive error with $\epsilon$ inverse polylogarithmic in $N$ using $\mathrm{polylog}(N)$ gates, this ends the proof.


\subsection{Importance sampling of decay rates}
\label{appendix:theorem3}

\begin{result}[Importance-sampling of decay rates] 
\label{res:decay}
    Under the conditions of Result~1, assume \mbox{$[h, X]=0$}. Then, there is an importance-sampling quantum algorithm that outputs an $\epsilon$-additive approximation of the decay rate $\nu_{kj} $ for the fermionic covariance matrix dynamics with probability equal to the spectral weight $|a_{kj}(0)|^2$. Each run requires $\mathrm{polylog}(N)$ elementary quantum gates.
\end{result}

We first note that, under the assumption $[h,X]=0$, the matrices $h$ and $X$ are simultaneously diagonalizable. We consider the eigen-decomposition of $X$ as
\begin{align}
    X\ket{\phi_k} = \nu_k \ket{\phi_k},
\end{align}
where $\{\ket{\phi_k}\}$ is an orthonormal eigenbasis of $X$ with associated eigenvalues $\{\nu_k\}$. Then,
\begin{align}
    \mathbb{X} := X \otimes I + I \otimes X
\end{align}
has eigenvectors $\ket{\phi_k}\otimes \ket{\phi_l}$ with eigenvalues $\nu_{kl} := \nu_k + \nu_l$. Expanding the input state in this basis gives: 
\begin{align}
    \ket{\psi(0)} = \ket{\mathrm{vec}(\Gamma(0))} = \sum_{k,l} a_{kl}(0)\, \ket{\phi_k}\otimes \ket{\phi_l}.
\end{align}

Using the efficient block-encoding of $X$ constructed for Result~1 in Section~\ref{app:fermions_block}, we can efficiently block-encode $\mathbb{X}$ by linear combination of unitaries. Hence, we can efficiently block-encode the unitary $U=e^{\i 2\pi \mathbb{X}/a_{\mathbb{X}}}$, with 
\begin{equation}
    a_{\mathbb{X}} \leq 2 a_X \leq 2s^2 = O(\mathrm{polylog}(N))
\end{equation} 
using the results of Section~\ref{app:fermions_block}. Now, we run a standard QPE algorithm on the initial state $\ket{\psi(0)}$. Let us expand on that for completeness. We introduce an $m$ qubit register, set $M= 2^m = \Theta(\alpha_{\mathbb{X}}/\epsilon)$, and define $\mathcal{U} := \sum_{r} \ketbra{r}{r} \otimes U^r$. The latter, using the standard decomposition in terms of powers of $2$ of $U$, can be implemented at a cost scaling as the cost of a single application of $U$ times $\Theta(\alpha_{\mathbb{X}}/\epsilon))$, resulting in a cost $\mathrm{polylog}(N)$ overall. Assuming for simplicity that all $\nu_k$ admit an $m$-bit representation
\begin{align}
    \ket{0^m}\ot \ket{\psi(0)} &= \sum_{k,l} a_{kl}(0) \ket{0^m,\phi_k,\phi_l} \xrightarrow{H^{\otimes m} \otimes I} \sum_{k,l,r} \frac{a_{kl}(0)}{\sqrt{M}} \ket{r,\phi_k,\phi_l} \\
    & \xrightarrow{\mathcal{U}} \sum_{k,l,r} \frac{a_{kl}(0)}{\sqrt{M}} e^{\i 2\pi r (\nu_k+\nu_l)/ a_{\mathbb{X}}} \ket{r,\phi_k,\phi_l} \\
    & \xrightarrow{\mathrm{QFT}^\dag \otimes I}
    \sum_{k,l,r,r'} \frac{a_{kl}(0)}{M} e^{\i 2\pi r\left[ (\nu_k+\nu_l)/ a_{\mathbb{X}} - \frac{r'}{M}\right]} \ket{r',\phi_k,\phi_l} 
    \\
    & =
    \sum_{k,l} a_{kl}(0) \ket{M(\nu_k + \nu_l)/ a_{\mathbb{X}}}\ot \ket{\phi_k}\ot\ket{\phi_l}. 
\end{align}
A measurement of the ancilla register then returns an integer representation of $\nu_k + \nu_l$ with probability $|a_{kl}(0)|^2$. Equivalently, each run of the algorithm samples the decay-rate distribution at resolution $\epsilon$: it first draws a decay channel with importance weight $|a_{kl}(0)|^2$, and then outputs an $\epsilon$-accurate estimate of the corresponding decay rate 

The case where $\nu_k$ does not admit an exact $m$-bit representation leads to the well-known and thoroughly studied problem of the bit discretization error~\cite{lin2022lecture},  which does not change the scaling presented here.


\subsection{Preparing the steady state covariance matrix}
\label{appendix:theorem4}

\begin{result}[Steady state]
\label{res:steady}
  Under the assumptions of Result~\ref{res:decay}, let $\Delta_{\mathcal{L}}$ be the spectral gap of the Lindbladian. Then, one can construct: (a) a unitary block-encoding $U_{\Gamma_{\infty}}$ of the steady state covariance matrix $\Gamma_\infty$ with an operator norm error $\epsilon$ and block-encoding scale-factor $\alpha_{\Gamma_{\infty}}= O( \Delta_{\mathcal{L}}^{-1}  \mathrm{polylog}(N))$; (b) a quantum state proportional to $\mathrm{vec}(\Gamma_\infty)$, with Euclidean norm error $\epsilon$. These constructions require $O \left (\Delta^{-1}_{\mathcal{L}} \mathrm{polylog}(N)\log (\Delta^{-1}_{\mathcal{L}} N/\epsilon) \right )$ elementary quantum gates. 
\end{result} 

Under the assumption that the dynamics is strictly dissipative we then have a unique asymptotic fixed point $\Gamma_\infty$ for the dynamics of the covariance matrix. This is determined from the matrix equation
\begin{equation}
    B \Gamma_{\infty} + \Gamma_{\infty} B^\top = - Y.
\end{equation}
We know that $\|B\| = \Theta(\|h\|, \|X\|) = \Theta(\mathrm{polylog}(N))$ and, from the part of the proof of Result~1 presented in Section~\ref{app:fermions_block}, we may assume access to a unitary block-encoding $U_B$ with a scale-factor $\alpha_B$. We assume it is asymptotically optimal, so that $\alpha_B = \Theta (\mathrm{polylog}(N))$, and indeed assume without loss of generality that $\alpha_B \ge 2 \|B\|$. We now rescale the problem to
\begin{equation}
    \tilde{B} \Gamma_{\infty} + \Gamma_{\infty} \tilde{B}^\top = - \tilde{Y},    
\end{equation}
where $\tilde{B} = B/ \alpha_B$, $\tilde{Y}= Y/\alpha_B$, and $\alpha_B \geq 2 \| B\|$. Hence, $\| \tilde{B}\| \leq 1/2$. Since $Y$ is at most $s^2$-sparse, its entries are $\Theta(1)$ and can be efficiently computed from the indices, an efficient block-encoding $U_Y$ of $Y$ can be constructed with $\alpha_Y = \Theta (\mathrm{polylog}(N))$ via standard sparse block-encoding methods. This, in turn, implies a block-encoding $U_{\tilde{Y}}$ of $\tilde{Y}$ with a scale-factor $\alpha_{\tilde{Y}} =\alpha_Y/\alpha_B = O (\mathrm{polylog}(N))$. Note that $\alpha_{\tilde{Y}}$ is an upper bound for $\|\tilde{Y}\|$. 

Next, consider $\tilde{\mathbb{B}} = \tilde{B} \otimes I + I \otimes \tilde{B}$. Since $[h,X]=0$, $\tilde{B}$ is normal. It then follows that $\tilde{\mathbb{B}}$ is normal. The eigenvalues of $\mathbb{B}$ are $-\nu_{i_1} - \nu_{i_2} + i( \varepsilon_{i_3} + \varepsilon_{i_4})$, where $\nu_j$ are the eigenvalues of $X$ and $i\varepsilon_k$ are the eigenvalues of $h$. Hence,
\begin{equation}
    \|\tilde{\mathbb{B}}^{-1}\| = O  \left(\frac{\alpha_B}{  \nu_{\mathrm{min}}}\right) = O\left(\frac{1}{ \nu_{\mathrm{min}}}\mathrm{polylog}(N) \right).
\end{equation}
The linear equation is defined on the space of covariance matrices $\Gamma$, however these have the constraint that $\Gamma^\top = -\Gamma$ and therefore $\ket{\mathrm{vec}(\Gamma)}$ lives in the anti-symmetric subspace spanned by $\frac{1}{\sqrt{2}}(\ket{\phi_j}\ket{\phi_k} - \ket{\phi_k}\ket{\phi_j})$. The spectral gap $\Delta_{\mathcal{L}}$ of the Lindbladian is defined as the minimum on the antisymmetric subspace of $\nu_{i_1} + \nu_{i_2}$. It follows that $\Delta_{\mathcal{L}} = \Omega(\nu_{\mathrm{min}})$. 
Hence, it follows that 
\begin{equation}
    \|\tilde{\mathbb{B}}^{-1}\| = O\left(\frac{1}{\Delta_\mathcal{L}}\mathrm{polylog}(N) \right).
\end{equation}

We now apply Theorem~1.1 of Ref.~\cite{somma2025quantum} for the case where $\tilde{\mathbb{B}}$ is normal to output the block-encoding $\Gamma_\infty/\alpha_{\Gamma_\infty}$. The query complexity to the block-encodings of $B$, its transpose and $Y$, as well as the elementary gate count complexity scale as
\begin{equation}
    O \left (\|\tilde{\mathbb{B}}^{-1}\| \log (\|\tilde{\mathbb{B}}^{-1}\| N/\epsilon) \right ) =O \left (\Delta^{-1}_{\mathcal{L}} \mathrm{polylog}(N)\log (\Delta^{-1}_{\mathcal{L}} N/\epsilon) \right ).
\end{equation}
The scale-factor $\alpha_{\Gamma_\infty}$ scales as
\begin{equation}
    \alpha_{\Gamma_\infty} = O \left ( \alpha_{\tilde{Y}}\|\tilde{\mathbb{B}}^{-1}\| \sqrt{\log (N/\epsilon)}  \right ) = O \left ( \Delta_{\mathcal{L}}^{-1} \mathrm{polylog}(N)  \right ).
\end{equation}

For the case where we want to output a vector proportional to $\mathrm{vec}(\Gamma_\infty)$, we instead consider the equation
\begin{align}
    \mathrm{vec}(\Gamma_\infty) = \mathbb{\tilde{B}}^{-1} \mathrm{vec}(\tilde{Y}).
\end{align}
We have already seen in Section~\ref{app:fermions_blockY} that a quantum state proportional to $\mathrm{vec}(\tilde{Y})$ can be prepared efficiently. Furthermore, the condition number of $\mathbb{\tilde{B}}$ is equal to that of $\mathbb{B}$, which reads
\begin{align}
    \kappa_{\mathbb{\tilde{B}}} = O\left(\frac{1}{\nu_{\mathrm{min}} } \mathrm{polylog}(N) \right).  
\end{align}
The complexity of outputting a vector proportional to $ \mathrm{vec}(\Gamma_\infty)$ with an optimal quantum linear solver is then~\cite{dalzell2024shortcut}:
\begin{equation}
    O(\alpha_{\mathbb{\tilde{B}}}  \kappa_{\mathbb{\tilde{B}}} \log(1/\epsilon)) = O\left(\frac{1}{\Delta_{\mathcal{L}} } \mathrm{polylog}(N) \right).
\end{equation}


\section{Nonlinear interaction picture example}
\label{app:nonlinear_ip}

Consider a $d$-dimensional system that unperturbed looks like $d$ uncoupled 1-dimensional systems:
\begin{equation}
    \dot{x} = A(x),\quad [A(x)]_i=a_i(x_i).
\end{equation}
Its Koopman modes are then known and given by
\begin{equation}
    \eta_i(x(t)) = \exp \left [ \lambda_i \int_{a_0}^{x_i(t)} \frac{dy}{a_i(y)} \right ],
\end{equation}
and evolve according to:
\begin{equation}
    \dot{\eta} = \Lambda \eta,
\end{equation}
where $\Lambda$ is diagonal with $\Lambda_{ii}=\lambda_i$. Now, let us add a perturbation, replacing $A$ with $F=A+\delta B$, so that the dynamics of $x$ is given by
\begin{equation}
    \dot{x} = A(x)  +\delta B(x).
\end{equation}
The evolution of the unperturbed Koopman modes after turning on the interaction becomes:
\begin{equation}
    \dot{\eta}_i = \lambda_i \eta_i+ [\delta B(x)]_i \frac{\partial \eta_i}{ \partial x_i}.
\end{equation}
Note that since we know expressions for modes, we also know their partial derivatives $\partial \eta_i/\partial x_i$, i.e., we have
\begin{equation}
    \frac{\partial \eta_i}{\partial x_i} a_i(x_i) = \lambda_i \eta_i \quad\Rightarrow\quad \frac{\partial \eta_i}{\partial x_i} =\frac{\lambda_i \eta_i}{a_i(x_i)}.
\end{equation}
Hence, we can write
\begin{equation}
    \dot{\eta} = \Lambda \eta + \delta\tilde{ B}(x),
\end{equation}
where $\delta\tilde{B}(x)$ is given by
\begin{equation}
    [\delta \tilde{B}(x)]_i  = [\delta B(x)]_i \frac{\lambda_i \eta_i}{a_i(x_i)}.
\end{equation}

Now, assuming that the interaction $\delta\tilde{B}(x)$ can be expanded in the powers of $\eta$,
\begin{equation}
    \delta\tilde{B}(x) = \sum_{k=0}^{\infty} B_k \eta^{\otimes k},
\end{equation}
we then arrive at the Carleman equation for the Koopman modes
\begin{equation}
    \dot{\eta} = B_0+(\Lambda+B_1) \eta + \sum_{k=2}^{\infty} B_k \eta^{\otimes k}.
\end{equation}
Note that the above assumption means that the original perturbation is given by
\begin{equation}
    [\delta B(x)]_i =  \frac{a_i(x_i)}{\lambda_i \eta_i} \left[\sum_{k=0}^{\infty} B_k \eta^{\otimes k}\right]_i,
\end{equation}
and so we are dealing with the following dynamical systems:
\begin{equation}
    \dot{x}_i = a_i(x_i)\left( 1 + \frac{1}{\lambda_i \eta_i} \left[\sum_{k=0}^{\infty} B_k \eta^{\otimes k}\right]_i\right).
\end{equation}


\section{Population dynamics example}
\label{app:karleman_k}


\subsection{Carleman expansion}

We start by slightly rewriting the original formulation, so that the problem can be treated using standard Carleman linearization technique. More precisely, the original system, when linearized, becomes unstable, as $\dot{x}_i=r_i x_i$ explodes to infinity for $r_i>0$. We thus perform a simple linear change of variables,
\begin{equation}
    y_i := 1- \frac{x_i}{X_i},
\end{equation}
so that the population dynamics is described by
\begin{equation}
    \label{eq:vac_dyn}
    \dot{y}_i = -r_iy_i(1-y_i)+ X_i (1-y_i)^2 \sum_{j,k} J_{i,jk} \eta_j \eta_k,\qquad \eta_j=\frac{y_j}{1-y_j}. 
\end{equation}
The new variables $y_i$ measure vacancy, i.e., what is the unused fraction of the unperturbed carrying capacity $X_i$ by population $x_i$ (with negative $y_i$ indicating the population is over its carrying capacity). 

Now, although the nonlinear term is not polynomial, we can perform a Taylor expansion around $y=0$ (corresponding to an expansion of the original populations around their unperturbed carrying capacities). More precisely, for $|y_i|<1$, we have
\begin{equation}
    \eta_j = \frac{y_j}{1-y_j} = \sum_{n=1}^{\infty} y_j^n.
\end{equation}
Moreover, since choosing Carleman truncation order $N_C$ disregards all polynomial terms of order higher than $N_C$, for a given $N_C$ the Taylor expansion can also be truncated at $N_C$. Hence, we are dealing with the following dynamical system:
\begin{equation}
    \label{eq:vac_dyn_carl}
    \dot{y}_i = -r_iy_i(1-y_i)+ X_i\sum_{j,k} J_{i,jk}\left(  \sum_{\substack{m,n\geq 1 \\ m+n\leq N_C}} y_j^m y_k^n -2y_i  \sum_{\substack{m,n\geq 1 \\ m+n\leq N_C-1}} y_j^m y_k^n+ y_i^2  \sum_{\substack{m,n\geq 1\\ m+n\leq N_C-2}}y_j^m y_k^n\right),
\end{equation}
which can be neatly rewritten as
\begin{equation}
    \dot{y} = \sum_{n=1}^{N_C} F_n y^{\otimes n},
\end{equation}
where the first five $F_n$ matrices are explicitly given by
\begin{aligns}
    [F_1]_{i,j} &= -r_i \delta_{ij}, \\
    [F_2]_{i,jk} &= r_i \delta_{ij}\delta_{ik}+X_iJ_{i,jk},\\
    [F_3]_{i,jkl} &= X_iJ_{i,jk}(-2\delta_{il}+\delta_{jl}+\delta_{kl}),\\
    [F_4]_{i,jklm} &= X_iJ_{i,jk}(\delta_{kl}\delta_{km}+\delta_{jl}\delta_{km}+\delta_{jl}\delta_{jm}-2\delta_{il}(\delta_{jm}+\delta_{km})+\delta_{il}\delta_{im}),\\
    [F_5]_{i,jklmn} &= X_iJ_{i,jk}((\delta_{kl}\delta_{km}\delta_{kn}+ \delta_{jl}\delta_{km}\delta_{kn}+ \delta_{jl}\delta_{jm}\delta_{kn}+ \delta_{jl}\delta_{jm}\delta_{jn})- 2\delta_{il}(\delta_{jm}\delta_{jn}\\&\quad+ \delta_{jm}\delta_{kn}+ \delta_{km}\delta_{kn})+ \delta_{il}\delta_{im}(\delta_{jn}+\delta_{kn})).
\end{aligns}

This system is then amenable to a standard Carleman linearization technique. Namely, introducing the Carleman vector
\begin{equation}
    g = [g^{(1)},g^{(2)},\dots,g^{(N_C)}],
\end{equation}
where each $g^{(k)}$ is $d^k$-dimensional and meant to approximate $y^{\ot k}$, and the upper-block-triangular Carleman matrix $C$ with blocks
\begin{equation}
    C_{i,i+j} = F_{j+1} \ot I^{\ot i-1} + I \ot F_{j+1} \ot I^{\ot i-2}+\dots + I^{\ot i-1}\ot F_{j+1}, 
\end{equation}
the Carleman-linearized version of the above problem is given by
\begin{equation}
    \dot{g}(t) = Cg(t),\qquad g(0)=[y(0),y(0)^{\ot2},\dots,y(0)^{\ot N_C}].
\end{equation}
The Carleman truncation error, for the chosen truncation order $N_C$, is then given by:
\begin{equation}
    \epsilon_C(t) = \|y(t)-g^{(1)}(t)\|,
\end{equation}
and when $\epsilon_C\xrightarrow{N_C\to\infty}0$, we say that the truncation error converges. Numerically, we say that it converges when we observe that $\epsilon_C$ decreases when $N_C$ increases. 


\subsection{Nonlinear interaction picture}

Instead of expanding Eq.~\eqref{eq:vac_dyn} around linear dissipative dynamics described by $\dot{y}_i=-r_iy_i$, we can alternatively expand it around the non-interacting quadratic systems $\dot{y}_i = -r_i y_i(1-y_i)$. The Koopman modes for such systems can be easily found and are given by
\begin{equation}
    \eta_i = \frac{y_i}{1-y_i},    
\end{equation}
with the corresponding eigenvalue $-r_i$. It is then straightforward to find the evolution equation for $\eta_i$ when the interaction is turned on:
\begin{equation}
\label{eq:G1G2}
    \dot{\eta}_i = -r_i \eta_i +X_i\sum_{jk}J_{i,jk} \eta_j \eta_k.
\end{equation}
The above can be concisely written as
\begin{equation}
    \dot{\eta}= G_1 \eta + G_2 \eta^{\ot 2},
\end{equation}
with
\begin{aligns}
    [G_1]_{i,j} &= -r_i \delta_{ij},\\
    [G_2]_{i,jk} &= X_i J_{i,jk}.
\end{aligns}

We can now use Carleman linearization for the evolution of the Koopman modes. Such nonlinear interaction picture method is analogous to the Carleman linearization described in the previous subsection. More precisely, we again introduce the vector $g$ and the Koopman-Carleman matrix $K$ given by $C$ but with $F_k$ matrices replaced by $G_k$, and solve $\dot{g}=Kg$ with the initial condition given by $g^{(k)}(0) = \eta^{\ot k}(0)$. The resulting truncation error, for the chosen truncation order $N_C$, is then given by:
\begin{equation}
    \epsilon_K(t) = \|y(t)-\tilde{y}(t)\|,
\end{equation}
where
\begin{equation}
    \tilde{y}_i(t) = \frac{g^{(1)}_i(t)}{1+g^{(1)}_i(t)}. 
\end{equation}
Note that the system is quadratic and dissipative, so that one can use analytic results specifying conditions for guaranteed error convergence~\cite{liu2021efficient,costa2023further,krovi2023improved,jennings2025quantum}. In particular, the truncation error $\epsilon_K$ is guaranteed to converge to zero with growing $N_C$ if $ R_{\mathrm{K}}<1$, where
\begin{equation}
    \label{eq:R_number}
    R_{\mathrm{K}} = \frac{\|G_2\|\|\eta(0)\|}{\min_i r_i}.
\end{equation}


\subsection{Proof of Result~2}

\setcounter{result}{1}
\begin{result}[Exponentially large nonlinear population dynamics -- formal]
    \label{res:populations_formal}
    Assume that $r_i>0$, $X_i>0$, and $J_{i,jk}\in \mathbb{R}$ from~Eq.~\eqref{Eq:populationdynamics} are efficiently computable from their indices, that $J$ is $\Theta(\mathrm{polylog}(d))$-sparse and we are given an efficient circuit preparing $\ket{\eta(0)}$. Then, if
    \begin{align}
    R_{\mathrm{K}} = \tfrac{\|G_2\|\|\eta(0)\|}{\min_i r_i} < 1,
    \end{align}
    where $[G_2]_{i,jk}= X_i J_{i,jk}$,
    there is a quantum algorithm outputting an $\epsilon$-accurate approximation to the history state $\propto \sum_{s=0}^{t/\Delta t} \| \eta(s \Delta t)\| \ket{\eta(s \Delta t)} \ket{s \Delta t}$ for time $t=\Theta(\mathrm{polylog}(d))$, target error $\epsilon=\Theta(1/\mathrm{polylog}(d))$, and timestep $\Delta t = O(1/\mathrm{polylog}(d))$.
\end{result}

We want to solve Eq.~\eqref{eq:G1G2} with $\eta(0) = \eta_0$. This can be done by applying the Carleman approach of Refs.~\cite{liu2021efficient, costa2023further} to the Koopman observables we introduced. We write a hierarchy of equations for $g=\{ \eta, \eta^{\otimes 2}, \eta^{\otimes 3}, \dots, \}$ and truncate at level $N_C$ by setting $\eta^{\otimes N_C+1} \equiv 0$.  This gives a set of ODE equations
\begin{align}
    \dot{g} = K g, \quad g(0) = (\eta_0, \eta_0^{\otimes 2}, \dots, \eta_0^{\otimes N_C}),
\end{align}
with solution $g$. 
Since $R_{\mathrm{K}} <1$, one can show there exists a rescaling $\gamma >0$ such that $\bar{\eta} = \eta/\gamma$ satisfies the rescaled equations
\begin{align}
    \dot{\bar{\eta}}= \bar{G}_1 \bar{\eta} + \bar{G}_2 \bar{\eta} \otimes \bar{\eta}, \quad \bar{\eta}(0) = \bar{\eta}_0/\gamma,
\end{align}
with $\bar{G}_2 = \gamma G_2$, $\bar{G}_1 = G_1$ and (1) $\|\bar{\eta}(0)\|<1$, (2) The rescaled Koopman matrix $\bar{K}$ has negative log-norm, \mbox{$\mu(\bar{K})<0$}~\cite{costa2023further, jennings2025quantum}. Assume this rescaling has been carried out, and drop the bar from now on.

From the above, one finds that the Koopman-Carleman hierarchy has exponential convergence, in the sense that if we take $N_C = O(\log(1/\epsilon))$ one has
\begin{align}
\label{eq:errorKoop}
\max_{\tau \in [0, t]}\| g(\tau)     - \tilde{g}(\tau)\| \leq \epsilon,
\end{align}
where $\tilde{g}(t) = (\eta(t), \eta(t)^{\otimes 2}, \dots, \eta(t)^{\otimes N_C})$, with $\eta(t)$ the exact solution to $\dot{\eta}= G_1 \eta + G_2 \eta \otimes \eta$ with $\eta(0) = \eta_0$~\cite{costa2023further, jennings2025quantum}.

The system of ODEs resulting from this nonlinear interaction picture can be solved via a quantum ODE solver. Under the assumptions of Result~\ref{res:populations_formal}, the matrices $G_1$, $G_2$ in Eq.~\eqref{eq:G1G2} can be efficiently block-encoded via sparse-access constructions~\cite{lin2022lecture}, with corresponding block-encoding prefactors $\alpha_i = O(\mathrm{polylog}(d))$, $i=1,2$. The Koopman-Carleman matrix $K$ can then be encoded via linear combination of unitaries with a block-encoding prefactor $O(N_C(\alpha_1 + \alpha_2))$~\cite{liu2021efficient}, with a circuit involving $\Theta(1)$ applications of the block-encodings of $G_1$, $G_2$. Hence, we obtain a block-encoding $K$ with prefactor $O(N_C \times \mathrm{polylog}(d)) = O(\mathrm{polylog}(d))$, since $N_C$ is a $O(\log \log(d))$ contribution. The cost of running the quantum ODE solver, in terms of number of times we need to apply the unitary block-encoding of $K$ and a unitary preparing the state $\ket{g(0)}$ is
\begin{align}
    Q=O\left( \alpha_K  \max_{\tau \in [0,t]} \|e^{K \tau}\|  t \ \log \frac{ t}{\epsilon } \right) = O(\mathrm{polylog}(d)),
\end{align}
since $\|e^{K \tau}\|  \leq 1$. 
(Note that, in fact, even better results exist in terms of the $t$ scaling~\cite{jennings2023cost, an2024fast}, but we will not need them here). The above translates into $O(\mathrm{polylog}(d))$ applications of the block-encodings of $G_1$, $G_2$, plus $O(\mathrm{polylog}(d)$ gates used in the circuit constructions, giving a total gate cost $O(\mathrm{polylog}(d))$. The state $\ket{g(0)}$ can be prepared with $O(N_C)= O(\log \log d)$ calls to the a unitary preparing $\ket{\eta(0)}$~\cite{liu2021efficient}, hence again $O(\mathrm{polylog}(d))$ gates in total. 

The algorithm outputs an $O(\epsilon)$-approximation to the history state
\begin{align}
    \propto \sum_{s=0}^{t/\Delta t} \| \tilde{g}(s \Delta t)\| \ket{\tilde{g}(s \Delta t)}_\infty \ket{s \Delta t}, 
\end{align}
where $\Delta t = O(1/\omega_K) = O(1/\mathrm{polylog}(d))$. From this,
the error bound~\eqref{eq:errorKoop}, and the fact that the amplitude concentrates in the $N_C=1$ block since $\| \eta(t)\| <1$ for all $t >0$, we can efficiently extract via amplitude amplification an $O(\epsilon)$-approximation to the history state
\begin{align}
    \propto \sum_{s=0}^{t/\Delta t} \| \eta(s \Delta t)\| \ket{\eta(s \Delta t)} \ket{s \Delta t}. 
\end{align}


\subsection{Simulation parameters, complexity of the model, and convergence guarantees}

\begin{figure}
    \centering
    \includegraphics[width=0.3\linewidth]
    {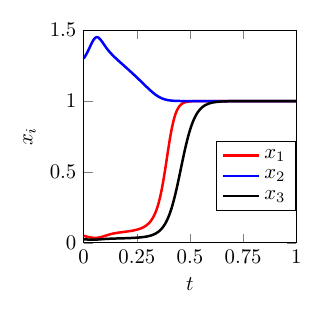}
    \includegraphics[width=0.3\linewidth]
    {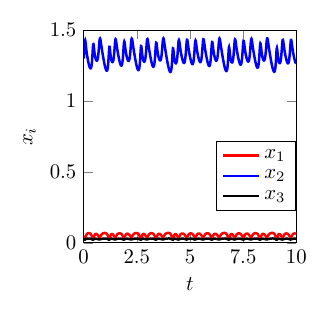}
    \includegraphics[width=0.3\linewidth]
    {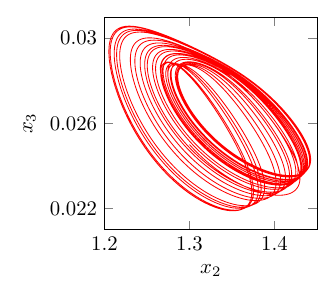}
    
    \caption{\textbf{Stable and chaotic evolution for population dynamics example.} The evolution of the considered population model for the initial condition $x(0)=[0.05,1.3,0.025]$ (left) and $x(0)=[0.048,1.3,0.025]$ (middle). While in the first case the system evolves to a stable equilibrium $(1,1,1)$, the second case exhibits a chaotic behavior. This can be seen looking at the projection onto the $x_2$-$x_3$ plane, where we see that the system's trajectory falls onto a strange attractor (right). }
    \label{fig:trajectories}
\end{figure}

For the numerical results presented in the main text, we chose the model with three populations and equal carrying capacities, $X_1=X_2=X_3=1$. The reproduction rates were chosen to be given by
\begin{equation}
    r_1 = 95.4912,\quad r_2=48.8281,\quad  r_3=30.1714,
\end{equation}
and the couplings read:
\begin{equation}
    J =\begin{pmatrix}
       -0.264803 & -13.6839 & 0.931878 & 0 & 983.541 & 69.1103 & 0 & 0 & 1.26601\\
       0.00120019 & -1.26625 & -0.00141069 & 0 & 46.6796 & 2.29013 & 0 & 0 & 0.000420895 \\
       -1.10445 & 42.4425 & 0.203853 & 0 & -477.852 & 17.3411 & 0 & 0 & 1.28499 
    \end{pmatrix}.
\end{equation}
This model is based on the dynamical system M5 found by the authors of Ref.~\cite{casas2016asymptotically}. It is the smallest possible (three-dimensional) system that is linearly stable with only negative real eigenvalues (i.e., there is a stable equilibrium sink) and exhibits chaotic behavior (in the form of a strange attractor). This choice was made deliberately to demonstrate that the considered model is rich enough to capture chaotic dynamics, see Fig.~\ref{fig:trajectories}. Finally, note that the $R$-number from Eq.~\eqref{eq:R_number} for the considered system is given by
\begin{equation}
     R_{\mathrm{K}} = 36.3316 \|\eta(0)\| = 36.3316 \sqrt{\sum_{i=1}^3 \left(\frac{1}{x_i(0)}-1\right)^2},
\end{equation}
and so we get a guaranteed error convergence for initial states satisfying
\begin{equation}
    \sum_{i=1}^3 \left(\frac{1}{x_i(0)}-1\right)^2 < 7.5758 \cdot 10^{-4}. 
\end{equation}
As is clear from the numerical results presented in the main text, the convergence region is in fact much larger.

\section{Large separation of Carleman parameters in monomial and Koopman coordinates}
\label{app:karleman_separation_r_numbers}

\setcounter{result}{6}
\begin{result}[Unbounded $R$--number separation]
    \label{res:R_separation}
    There exists a family of quadratic nonlinear systems, parametrized by $\beta>1$, such that the $R$-number for the Carleman system $R_C \xrightarrow{\beta\to\infty} \infty$, whereas the $R$-number for the nonlinear interaction picture system $R_K \xrightarrow{\beta\to\infty} 0$.
\end{result}

In this section, we construct an explicit nonlinear system for which the Carleman parameter $R$ depends strongly on the coordinate representation of the dynamics. In particular, we exhibit a system whose dynamics in the original variables $x$ have $R_x \gg 1$, while after transforming to variables $\eta$ aligned with the Koopman eigenmodes of the unperturbed dynamics, the resulting system satisfies $R_\eta \ll 1$.

To highlight the role of Koopman eigenmodes, we begin with a system written as a perturbation of an integrable nonlinear flow
\begin{align}
    \dot{x} = A(x) + \delta B(x),
\end{align}
where $\delta \ge 0$ is small. The variables $\eta$ will be defined as Koopman eigenmodes of the unperturbed system
\begin{align}
    \dot{x} = A(x).
\end{align}
We then derive the dynamical equations for $\eta$ in the perturbed system and analyze the corresponding Carleman parameters.


\subsection{Quadratic representation of the dynamics}

Consider the $d$-dimensional nonlinear ODE
\begin{align}
\dot{x}=Fx+v-x(c^\dagger x+\alpha),
\end{align}
where the coefficients $F \in \mathbb{C}^{d \times d}$, $v,c \in \mathbb{C}^d$ and $\alpha \in \mathbb{C}$ depend smoothly on a parameter $\delta$. The system can be viewed as
\begin{align}
\dot{x}=A(x)+\delta B(x),
\end{align}
with the unperturbed dynamics corresponding to $\delta=0$.

Following the standard lifting construction, we introduce variables $u \in \mathbb{C}^d$ and $w \in \mathbb{C}$ such that
\begin{align}
x=\frac{u}{w}.
\end{align}
The lifted variables obey a linear system
\begin{align}
\frac{d}{dt}
\begin{pmatrix}
u\\
w
\end{pmatrix}
=
H_x
\begin{pmatrix}
u\\
w
\end{pmatrix},
\qquad
H_x=
\begin{pmatrix}
F & v\\
c^\dagger & \alpha
\end{pmatrix}.
\end{align}
This representation allows nonlinear coordinate transformations in $x$ to be implemented as linear transformations in the lifted space.


\subsection{Nonlinear change of variables}

We now introduce the nonlinear change of variables that will later be chosen
so that the resulting variables coincide with Koopman eigenmodes of the
unperturbed system:
\begin{align}
\eta=\frac{Ax}{b^\dagger x+1},
\end{align}
where $A \in \mathbb{C}^{d \times d}$ is invertible and $b\in\mathbb{C}^d$ is arbitrary.

In the lifted variables, this transformation corresponds to
\begin{align}
\begin{pmatrix}
\tilde{u}\\
\tilde{w}
\end{pmatrix}
=
P
\begin{pmatrix}
u\\
w
\end{pmatrix},
\qquad
P=
\begin{pmatrix}
A & 0\\
b^\dagger & 1
\end{pmatrix}.
\end{align}
The new coordinates are
\begin{align}
\eta=\frac{\tilde{u}}{\tilde{w}}.
\end{align}

Under this transformation the lifted dynamics become
\begin{align}
\frac{d}{dt}
\begin{pmatrix}
\tilde{u}\\
\tilde{w}
\end{pmatrix}
=
H_\eta
\begin{pmatrix}
\tilde{u}\\
\tilde{w}
\end{pmatrix},
\qquad
H_\eta=PH_xP^{-1}.
\end{align}
The inverse of $P$ is
\begin{align}
    P^{-1}=\begin{pmatrix}
        A^{-1} & 0 \\
        -b^\dagger A^{-1} & 1
    \end{pmatrix}.
\end{align}
Therefore, by a direct calculation, 
\begin{align}
    H_\eta=\begin{pmatrix}
        \tilde{F} & \tilde{v} \\
        \tilde{c}^\dagger & \tilde{\alpha}
    \end{pmatrix},
\end{align}
where
\begin{gather}
\tilde{F}=A(F-vb^\dagger)A^{-1}, \qquad
\tilde{v}=Av, \\
\tilde{c}^\dagger=
\left(b^\dagger F+c^\dagger-(b^\dagger v+\alpha)b^\dagger\right)A^{-1}, \qquad
\tilde{\alpha}=b^\dagger v+\alpha.
\end{gather}

Projecting back to nonlinear variables yields
\begin{align}
\dot{\eta}=\tilde{F}\eta+\tilde{v}-\eta(\tilde{c}^\dagger \eta+\tilde{\alpha}).
\end{align}

Thus both coordinate systems lead to quadratic nonlinear ODEs.


\subsection{Koopman structure of the unperturbed dynamics}

We now specify the parameters so that the variables $\eta$ become Koopman eigenmodes of the unperturbed system. Let $\gamma>\beta\gg1$, let $A$ be unitary, and define
\begin{align}
b=\beta A^{-1}e_d .
\end{align}
For any $\delta \in [0, 1)$, let
\begin{align}
D(\delta)=\operatorname{diag}(1-\delta,\lambda_2,\lambda_3,\dots,\lambda_{d-1},1-\gamma),    
\end{align}
where $\lambda_2, \lambda_3,\dots, \lambda_{d-1} \in \mathbb{R}$ satisfy $\lambda_j \le 1-2\beta$ for $j=2,3,\dots,d-1$. Then define
\begin{gather}
F(\delta)=A^{-1}D(\delta)A, \qquad
v(\delta)=\delta A^{-1}e_d, \\
c(\delta)=-F(\delta)^\dagger b+(b^T v^*(\delta)+1)b+\delta A^{-1}e_1, \qquad
\alpha(\delta)=1.
\end{gather}

A direct calculation yields
\begin{gather}
\tilde{F}(\delta)=D(\delta)-\beta \delta e_d e_d^\dagger, \qquad \tilde{v}(\delta)=\delta e_d ,\\
\tilde{c}(\delta)
=\delta e_1, \qquad \tilde{\alpha}(\delta)=1+\beta \delta.
\end{gather}
In particular, when $\delta=0$, we have
$\tilde{F}(0)=D(0)$, $\tilde v(0)=\tilde{c}(0)=0$, and $\tilde{\alpha}(0)=1$. Thus, in this case, both the driving term and the quadratic interaction term vanish in the $\eta$ variables, and the system becomes
\begin{align}
\dot{\eta}
=
\operatorname{diag}(0,\lambda_2-1,\lambda_3-1,\dots,\lambda_{d-1}-1,-\gamma)\,\eta .
\end{align}
Thus each component $\eta_j$ evolves independently,
\begin{align}
\dot{\eta}_j=\lambda_j' \eta_j
\end{align}
for suitable $\lambda_j'$, showing that the variables $\eta_j$ are Koopman eigenmodes of the unperturbed system.


\subsection{Large separation of \texorpdfstring{$R_x$ and $R_\eta$}{R-numbers}}

Consider a quadratic ODE system
\begin{align}
\dot{z}=F_0+F_1z+F_2z^{\otimes2},\qquad z(0)=z_0.
\end{align}
We define
\begin{align}
R \defeq
\frac{\|F_2\|\|z_0\|+\dfrac{\|F_0\|}{\|z_0\|}}
{|\mu(F_1)|},
\end{align}
where
$\mu(F_1)=\lambda_{\max}\!\left(\frac{F_1+F_1^\dagger}{2}\right)$ is the logarithmic norm of $F_1$.

For the $x$-system
\begin{align}
\dot{x}=v+(F-\alpha I)x-(I\otimes c^\dagger)x^{\otimes2},
\end{align}
we therefore identify
$F_0=v$, $F_1=F-\alpha I$, and $F_2=-I\otimes c^\dagger$.
Similarly, for the $\eta$-system
\begin{align}
\dot{\eta}=\tilde v+(\tilde F-\tilde\alpha I)\eta-(I\otimes\tilde c^\dagger)\eta^{\otimes2},
\end{align}
the coefficients are
$\tilde F_0=\tilde v$, $\tilde F_1=\tilde F-\tilde\alpha I$, and $\tilde F_2=-I\otimes\tilde c^\dagger$.

From the construction, we have $v(\delta)=\delta A^{-1}e_d$ and
$\tilde v(\delta)=\delta e_d$,  and hence $\|F_0(\delta)\|=\|\tilde F_0(\delta)\|=\delta$. The quadratic coefficients satisfy
\begin{align}
c(\delta)=\delta A^{-1}e_1+(\gamma+\beta\delta)\beta A^{-1}e_d,
\end{align}
and therefore
\begin{align}
\|F_2(\delta)\|=\|c(\delta)\|
=\sqrt{\delta^2+(\gamma+\beta\delta)^2\beta^2}
\approx (\gamma+\beta\delta)\beta .
\end{align}
By contrast, $\tilde c(\delta)=\delta e_1$, and
$\|\tilde F_2(\delta)\|=\delta$. The logarithmic norms satisfy
\begin{align}
\mu(F_1(\delta))=-\delta, \qquad \mu(\tilde F_1(\delta))=-(\beta+1)\delta.     
\end{align}
Finally, choosing $x_0=A^{-1}e_1$ gives $\eta_0=e_1$ and hence $\|x_0\|=\|\eta_0\|=1$. 

It follows that for any fixed $\delta \in (0,1)$,
\begin{align}
R_x
=\frac{\delta+\sqrt{\delta^2+(\gamma+\beta\delta)^2\beta^2}}{\delta}
\ge
\frac{\delta+(\gamma+\beta\delta)\beta}{\delta}
=
\frac{\gamma\beta}{\delta}+\beta^2+1
\gg 1,
\end{align}
while
\begin{align}
R_\eta
=
\frac{2}{\beta+1}
\ll1 .
\end{align}
Thus, by taking $\beta$ sufficiently large and $\gamma>\beta$, the separation between $R_x$ and $R_\eta$ can be made arbitrarily large.

In this example, the same dynamical system admits two quadratic representations with markedly different Carleman parameters. In the original coordinates $x$, the parameter $R_x$ is extremely large, while in the Koopman-adapted coordinates $\eta$, where the unperturbed dynamics are diagonalized, the parameter $R_\eta$ becomes small. This example illustrates that aligning coordinates with Koopman eigenmodes can significantly simplify the effective nonlinear structure of the system and reduce the $R$-number.

The construction above provides a concrete instance of a more general principle: the apparent strength of nonlinearity is highly coordinate-dependent. In particular, coordinates aligned with Koopman eigenmodes can place the dynamics in a regime where Carleman linearization becomes efficient. 

The above construction is conceptually related to normal form transformations in dynamical systems, such as those arising in the Poincar\'{e}–Dulac theorem, where nonlinear changes of variables are used to simplify the structure of the dynamics by eliminating non-essential nonlinear terms (see~\cite{jennings2025quantum} for a relevant discussion). In contrast to the local and perturbative nature of Poincar\'{e}–Dulac normal forms, our transformation is global and explicitly constructed to align with Koopman eigenmodes of the unperturbed system. This alignment leads to a significant reduction in the effective nonlinearity, as quantified by the $R$-number.


\section{Spectral Quantum Koopman Algorithm for Oscillatory Modes}
\label{app:spectralqka}

In this section, we present a Spectral Quantum Koopman Algorithm (Spectral-QKA) for extracting the eigenfrequencies of oscillatory Koopman modes governing the late-time dynamics of observables. Here is a more formal statement compared to the one given in the main text:

\setcounter{result}{2}
\begin{result}[Importance sampling of oscillatory modes -- formal]
\label{res:spectral}
Assume access to a block-encoding of a normal Koopman matrix $K$ with scale factor $\alpha$, and a unitary procedure to prepare $\ket{g(0)}$ as in Eq.~\eqref{eq:initialdataspectral}, with
\[
\frac{\sum_{k:\mathrm{Re}(\lambda_k)=0}|a_k(0)|^2}
{\sum_k |a_k(0)|^2}
=
\Omega(1).
\]
Then the algorithm outputs a random variable $X$ whose distribution is $O(\zeta)$-close in total variation distance to an ideal random variable $Y$ defined as follows: first sample an oscillatory mode $k$ with probability
\[
\frac{|a_k(0)|^2}
{\sum_{j:\mathrm{Re}(\lambda_j)=0}|a_j(0)|^2},
\]
and then output a random variable $Y_k$ that is $\epsilon$-close to
$\mathrm{Im}(\lambda_k)$ with probability at least $1-\delta$. The query complexity is
\begin{align}
\mytO{
    \alpha
    \left(
        \frac{1}{\Delta}\log\frac{1}{\zeta}
        +
        \frac{1}{\epsilon}\log\frac{1}{\delta}
    \right)
    \mylogpow{2}{\frac{1}{\zeta}}
},
\end{align}
where
\[
\Delta
=
-\max_{k:\mathrm{Re}(\lambda_k)<0}
\mathrm{Re}(\lambda_k)
\]
is the spectral gap.
\end{result}

A key challenge in this setting is that standard quantum phase estimation (QPE) produces Dirichlet-type distributions with only polynomial decay when the target frequencies are not aligned with the discrete Fourier grid, leading to poor concentration of measurement outcomes. To overcome this limitation, we employ a windowed QPE procedure based on the Kaiser window, which achieves exponential tail decay and yields sharply concentrated frequency estimates without requiring any grid-alignment assumptions.

The main algorithmic ingredient is the preparation of a nonuniform history state
\begin{equation}
    \sum_{j=0}^{J-1}
    \beta_j
    \ket{j}
    \ket{g(T_1+j\Delta t)},
\end{equation}
where the coefficients $\beta_j$ are chosen according to the Kaiser window. We prepare this state using a linear-system embedding of the ODE together with a uniform-size construction for the different target times. More generally, the same construction can be used to prepare weighted history states with other efficiently preparable coefficients, and may therefore be useful beyond the spectral-estimation setting considered here.

Combining this nonuniform history-state preparation procedure with the Kaiser-window Fourier transform enables efficient extraction of oscillatory mode frequencies in the late-time regime. It would be of interest to extend these ideas to more general settings, in particular to pseudo-spectral Quantum Koopman frameworks~\cite{yang2026towards}, which may provide a robust approach to handling the spectral properties of non-normal Koopman generators.

\subsection{Setup}

Consider the Koopman evolution
\begin{equation}
    \dot{g}(t)=K g(t),
    \qquad
    g(t)=e^{Kt}g(0),
\end{equation}
where $K$ is restricted to a finite-dimensional approximately invariant subspace.

We assume that $K$ is normal, and hence unitarily diagonalizable:
\begin{equation}
    K = V \Lambda V^\dagger,
    \qquad
    \Lambda = \diag{\lambda_k},
    \qquad
    \lambda_k = -\mu_k + \i \omega_k,
\end{equation}
where $V$ is unitary and $\mu_k \ge 0$. Let $\{\ket{\phi_k}\}$ denote the
orthonormal eigenbasis of $K$, given by the columns of $V$.

We partition the spectrum into oscillatory and decaying modes:
\begin{equation}
    S := \{k : \mu_k = 0\},
    \qquad
    F := \{k : \mu_k \ge \Delta\},
\end{equation}
for some spectral gap $\Delta>0$. Define the corresponding orthogonal
projectors
\begin{equation}
    \Pi_S
    \defeq
    \sum_{k\in S}\ket{\phi_k}\bra{\phi_k},
    \qquad
    \Pi_F
    \defeq
    \sum_{k\in F}\ket{\phi_k}\bra{\phi_k}.
\end{equation}
Then $\Pi_S+\Pi_F=I$, and  $\Pi_S\Pi_F=0$.

The initial condition admits the expansion
\begin{equation}
    \ket{g(0)}
    =
    \sum_k a_k\ket{\phi_k},
\end{equation}
and hence
\begin{align}
    \Pi_S\ket{g(0)}
    =
    \sum_{k\in S}a_k\ket{\phi_k},\qquad
    \Pi_F\ket{g(0)}
    =
    \sum_{k\in F}a_k\ket{\phi_k}.
\end{align}
Without loss of generality, we assume $\|g(0)\|=1$. Define
\begin{equation}
    \|a_S\|^2
    \defeq
    \|\Pi_S g(0)\|^2
    =
    \sum_{k\in S}|a_k|^2,
\end{equation}
and assume
\begin{equation}
    \|a_S\|^2\ge\eta>0.
\end{equation}

For any $t\ge0$,
\begin{equation}
    \Pi_S\ket{g(t)}
    =
    \sum_{k\in S}
    a_k e^{\i\omega_k t}\ket{\phi_k},
\end{equation}
and therefore
\begin{equation}
    \|\Pi_S g(t)\|
    =
    \|a_S\|
    \ge
    \sqrt{\eta}.
\end{equation}

We assume access to a block-encoding $U_K$ of $K$ with scale factor $\alpha$, and a unitary $U_0$ preparing $\ket{g(0)}$.

For convenience, we write $\bar{v}=v/\|v\|$ to denote the normalized version of a nonzero vector $v$ throughout this section.

\subsection{Suppression of decaying modes}

Let $\epsilon_1\in(0,\sqrt{\eta})$ and define
\begin{equation}
    T_1
    \defeq
    \frac{\ln(1/\epsilon_1)}{\Delta}.
\end{equation}
For all $t\ge T_1$,
\begin{align}
    \|g(t)-\Pi_S g(t)\|^2
    =
    \|\Pi_F g(t)\|^2
    =
    \sum_{k\in F}
    |a_k|^2 e^{-2\mu_k t}
    \le
    e^{-2\Delta t}
    \le
    \epsilon_1^2.
\end{align}
Thus,
\begin{equation}
    \|g(t)-\Pi_S g(t)\|
    \le
    \epsilon_1,
    \qquad
    \forall t\ge T_1.
\end{equation}

Thus, after time $T_1$, the decaying component of the Koopman evolution is uniformly suppressed. This will allow the nonuniform history state prepared over times $t\ge T_1$ to be approximated by a history state supported entirely on the oscillatory modes.

\subsection{Windowed phase estimation}

Let $\{\beta_j\}_{j=0}^{J-1}$ satisfy
$\sum_{j=0}^{J-1}|\beta_j|^2=1$.
For a unitary $U$ with eigenpair
\begin{equation}
    U\ket{\phi}
    =
    e^{\i\theta}\ket{\phi},
    \qquad
    \theta\in[-\pi,\pi),
\end{equation}
consider the windowed phase-estimation procedure
\begin{equation}
    \sum_{j=0}^{J-1}
    \beta_j\ket{j}\ket{\phi}
    \mapsto
    \sum_{j=0}^{J-1}
    \beta_j e^{\i j\theta}
    \ket{j}\ket{\phi},
\end{equation}
followed by the inverse quantum Fourier transform on the first register. The resulting state is
\begin{equation}
    \sum_{\ell=0}^{J-1}
    \gamma_\ell(\theta)
    \ket{\ell}\ket{\phi},
\end{equation}
where
\begin{equation}
    \gamma_\ell(\theta)
    \defeq
    \frac{1}{\sqrt{J}}
    \sum_{j=0}^{J-1}
    \beta_j
    e^{\i j(\theta-2\pi\ell/J)}.
\end{equation}

We decode the measurement outcome as a signed phase:
\begin{equation}
    \hat{\theta}_\ell
    \defeq
    \begin{cases}
        2\pi\ell/J,
        & 0\le \ell<J/2,\\[1mm]
        2\pi\ell/J-2\pi,
        & J/2\le \ell\le J-1.
    \end{cases}
\end{equation}
Thus, outcomes with $\ell\ge J/2$ are interpreted as negative phases. This is the natural real-valued decoding when the phase is represented in the principal interval $[-\pi,\pi)$.

Since phase estimation is periodic, ordinary real-valued accuracy is ambiguous near the branch cut at $\pm\pi$. We therefore assume a Nyquist-margin condition
\begin{equation}
    \theta\in[-\pi+c,\pi-c],
\end{equation}
where $c>0$ is a fixed constant, e.g., $c=0.1\pi$. For target accuracy
$0<\varepsilon<c$, the Kaiser-window analysis in Appendix~D of
Ref.~\cite{berry2024analyzing} implies that, for sufficiently small
$\varepsilon,\delta>0$, there exist coefficients
$\{\beta_j\}_{j=0}^{J-1}$, for an odd integer $J$ satisfying
\begin{equation}
    J
    =
    \Theta\!\left(
        \frac{\log(1/\delta)}{\varepsilon}
    \right),
\end{equation}
such that, uniformly for all
$\theta\in[-\pi+c,\pi-c]$,
\begin{equation}
    \sum_{\ell:\,
    |\hat{\theta}_\ell-\theta|>\varepsilon}
    |\gamma_\ell(\theta)|^2
    \le
    \delta.
\end{equation}
That is, the decoded phase-estimation outcome is $\varepsilon$-accurate as an ordinary real-valued estimate with probability at least $1-\delta$.

One convenient choice of coefficients $\{\beta_j\}_{j=0}^{J-1}$ with this scaling is the shifted discretized Kaiser window~\cite{berry2024analyzing}
\begin{equation}
\label{eq:kaiser-window}
    \beta_j
    :=
    \frac{1}{\sqrt{C}}
    \frac{
        I_0\!\left(
            \pi\sigma
            \sqrt{
                1-
                \left(
                    \frac{2j-(J-1)}{J-1}
                \right)^2
            }
        \right)
    }{
        I_0(\pi\sigma)
    },
    \qquad
    j=0,1,\dots,J-1,
\end{equation}
where $I_0$ denotes the modified Bessel function of the first kind and $C$ is a normalization constant ensuring
$\sum_j|\beta_j|^2=1$. The parameter $\sigma>0$ controls the trade-off between main-lobe width and tail suppression. With an appropriate choice
\begin{equation}
    \sigma
    =
    \Theta(\log(1/\delta)),
\end{equation}
this window yields exponentially decaying tails, giving the stated scaling. We refer to this variant of quantum phase estimation as \emph{Kaiser-QPE}.

\subsection{Algorithm}

Let $\Delta t>0$, and define $T_2=(J-1)\Delta t$ and
$T=T_1+T_2$. For $j=0,1,\dots,J-1$, define
$t_j \defeq T_1+j\Delta t$.

Let $\{\beta_j\}_{j=0}^{J-1}$ be the Kaiser-window coefficients defined in
Eq.~\eqref{eq:kaiser-window}, and define the unnormalized nonuniform history state
\begin{equation}
\label{eq:spectral-history-state}
    \ket{H}
    \defeq
    \sum_{j=0}^{J-1}
    \beta_j
    \ket{j}
    \ket{g(t_j)}.
\end{equation}

The algorithm proceeds as follows.

\begin{enumerate}
    \item
    Using the procedure described in Appendix~\ref{app:nonuniform-history-state},
    prepare a state $\ket{\widehat{H}}$ satisfying
    \begin{equation}
        \|
            \ket{\widehat{H}}
            -
            \ket{\bar{H}}
        \|
        \le
        \epsilon_{\rm hist}.
        \label{eq:history-state-error}
    \end{equation}

    \item
    Apply the inverse quantum Fourier transform to the first register.

    \item
    Measure the first register. If the outcome is
    $\ell\in\{0,1,\dots,J-1\}$, output
    \begin{equation}
        \hat{\omega}_\ell
        \defeq
        \begin{cases}
            \dfrac{2\pi\ell}{J\Delta t},
            & 0\le\ell\le\dfrac{J-1}{2},
            \\[2mm]
            \dfrac{2\pi\ell}{J\Delta t}
            -
            \dfrac{2\pi}{\Delta t},
            &
            \dfrac{J+1}{2}
            \le
            \ell
            \le
            J-1.
        \end{cases}
    \end{equation}
\end{enumerate}

The decoding rule above is the frequency version of the signed phase decoding
from the previous subsection. The Nyquist-margin assumption becomes
\begin{equation}
    -\pi+c
    \le
    \omega_k\Delta t
    \le
    \pi-c
\end{equation}
for every relevant frequency $\omega_k$. Under this condition, a phase
accuracy $\varepsilon<c$ corresponds to frequency accuracy
\begin{equation}
    |\hat{\omega}_\ell-\omega_k|
    \le
    \frac{\varepsilon}{\Delta t}.
\end{equation}

\subsection{Output distribution}

Define the oscillatory component of the history state by
\begin{equation}
    \ket{H_S}
    \defeq
    (I\otimes \Pi_S)\ket{H}
    =
    \sum_{j=0}^{J-1}
    \beta_j
    \ket{j}
    \Pi_S\ket{g(t_j)}.
\end{equation}
Since
$\sum_j|\beta_j|^2=1$ and
$\|\Pi_S g(t_j)\|=\|a_S\|$ for every $j$, we have
\begin{equation}
    \|\ket{H_S}\|
    =
    \|a_S\|
    \ge
    \sqrt{\eta}.
\end{equation}

Moreover, since $t_j\ge T_1$ for every $j$,
\begin{align}
    \|
        \ket{H}
        -
        \ket{H_S}
    \|^2
    =
    \sum_{j=0}^{J-1}
    |\beta_j|^2
    \|
        g(t_j)-\Pi_S g(t_j)
    \|^2
    \le
    \epsilon_1^2.
\end{align}
Therefore,
\begin{align}
    \|
        \ket{\bar H}
        -
        \ket{\bar H_S}
    \|
    \le
    \frac{
        2
        \|
            \ket H-\ket{H_S}
        \|
    }{
        \|\ket{H_S}\|
    }
    \le
    \frac{2\epsilon_1}{\sqrt{\eta}}.
    \label{eq:history-oscillatory-approx}
\end{align}

Combining Eqs.~\eqref{eq:history-state-error} and
\eqref{eq:history-oscillatory-approx}, the state prepared before the inverse Fourier transform satisfies
\begin{equation}
    \|
        \ket{\widehat H}
        -
        \ket{\bar H_S}
    \|
    =
    O\!\left(
        \frac{\epsilon_1}{\sqrt{\eta}}
        +
        \epsilon_{\rm hist}
    \right).
    \label{eq:spectral-state-total-error}
\end{equation}

We now expand the ideal oscillatory history state. Since
\begin{equation}
    \Pi_S\ket{g(t)}
    =
    \sum_{k\in S}
    a_k
    e^{\i\omega_k t}
    \ket{\phi_k},
\end{equation}
we have
\begin{align}
    \ket{\bar H_S}
    &=
    \frac{1}{\|a_S\|}
    \sum_{j=0}^{J-1}
    \beta_j
    \ket j
    \Pi_S\ket{g(T_1+j\Delta t)}
    \\
    &=
    \frac{1}{\|a_S\|}
    \sum_{k\in S}
    a_k
    e^{\i\omega_k T_1}
    \left(
        \sum_{j=0}^{J-1}
        \beta_j
        e^{\i\omega_k j\Delta t}
        \ket j
    \right)
    \ket{\phi_k}.
\end{align}

After applying the inverse quantum Fourier transform to the first register, let $p$ denote the actual measurement distribution and let $\tilde p$ denote the ideal distribution
\begin{equation}
\label{eq:spectralqka-ideal-distribution}
    \tilde p(\ell)
    =
    \sum_{k\in S}
    \frac{|a_k|^2}{\|a_S\|^2}
    |\gamma_\ell(\omega_k\Delta t)|^2.
\end{equation}
Equation~\eqref{eq:spectral-state-total-error} implies
\begin{equation}
\label{eq:spectralqka-tv}
    \|p-\tilde p\|_{\rm TV}
    =
    O\!\left(
        \frac{\epsilon_1}{\sqrt{\eta}}
        +
        \epsilon_{\rm hist}
    \right).
\end{equation}

Thus, the output distribution is close, in total variation distance, to a mixture of Kaiser-window phase-estimation distributions centered at the oscillatory Koopman frequencies, with weights proportional to $|a_k|^2$.

\subsection{Frequency estimation guarantee}

Let $\tilde p$ denote the ideal distribution in
Eq.~\eqref{eq:spectralqka-ideal-distribution}, and let $p$ denote the actual output distribution. By Eq.~\eqref{eq:spectralqka-tv},
\begin{equation}
    \|p-\tilde p\|_{\rm TV}
    \le
    \varepsilon_{\rm tot},
    \qquad
    \varepsilon_{\rm tot}
    :=
    O\!\left(
        \frac{\epsilon_1}{\sqrt{\eta}}
        +
        \epsilon_{\rm hist}
    \right).
\end{equation}

Fix $k\in S$, and define the event
\begin{equation}
    E_k
    =
    \{
        \ell:
        |\hat{\omega}_\ell-\omega_k|
        \le
        \epsilon
    \}.
\end{equation}
By the guarantee of Kaiser-QPE applied to the $k$th eigencomponent,
\begin{equation}
    \sum_{\ell\in E_k}
    |\gamma_\ell(\omega_k\Delta t)|^2
    \ge
    1-\delta.
\end{equation}
Consequently,
\begin{equation}
    \tilde p(E_k)
    \ge
    (1-\delta)
    \frac{|a_k|^2}{\|a_S\|^2}.
\end{equation}
Using the total variation bound,
\begin{align}
    p(E_k)
    \ge
    \tilde p(E_k)
    -
    \|p-\tilde p\|_{\rm TV}
    \ge
    (1-\delta)
    \frac{|a_k|^2}{\|a_S\|^2}
    -
    \varepsilon_{\rm tot}.
\end{align}

Thus, the algorithm outputs an $\epsilon$-accurate estimate of $\omega_k$ with probability at least
\begin{equation}
    (1-\delta)
    \frac{|a_k|^2}{\|a_S\|^2}
    -
    O\!\left(
        \frac{\epsilon_1}{\sqrt{\eta}}
        +
        \epsilon_{\rm hist}
    \right).
\end{equation}

Equivalently, up to total variation error
$O(\epsilon_1/\sqrt{\eta}+\epsilon_{\rm hist})$, the output distribution can be generated by first sampling an oscillatory mode $k$ with probability
$\frac{|a_k|^2}{\|a_S\|^2}$,
and then sampling from the corresponding Kaiser-QPE distribution centered at
$\omega_k$.

\subsection{Parameter selection and overall complexity}

Assume that the relevant oscillatory frequencies satisfy $|\omega_k| \le
    \omega_{\max}$, for all $k\in S$.
Since $\omega_{\max}
    \le \|K\|
    \le \alpha$,
we may choose $\Delta t$ so that
\begin{equation}
    \omega_{\max}\Delta t
    \le
    \pi-c
\end{equation}
for some fixed constant $c>0$. Thus,
\begin{equation}
    \Delta t
    =
    \Theta\!\left(
        \frac{1}{\omega_{\max}}
    \right).
\end{equation}
Under this choice, all relevant phases
\begin{equation}
    \theta_k
    =
    \omega_k\Delta t
\end{equation}
lie in the interval
$[-\pi+c,\pi-c]$, so the signed decoding in Kaiser-QPE gives an ordinary real-valued frequency estimate without branch-cut ambiguity.

Since the required phase-estimation accuracy is
\begin{equation}
    \varepsilon
    =
    \epsilon\Delta t,
\end{equation}
the Kaiser-QPE construction requires
\begin{align}
    J
    =
    \Theta\!\left(
        \frac{\log(1/\delta)}
        {\epsilon\Delta t}
    \right)
    =
    \Theta\!\left(
        \frac{
            \omega_{\max}
            \log(1/\delta)
        }{
            \epsilon
        }
    \right).
\end{align}
Consequently,
\begin{equation}
    T_2
    =
    (J-1)\Delta t
    =
    \Theta\!\left(
        \frac{\log(1/\delta)}{\epsilon}
    \right),
\end{equation}
while
\begin{equation}
    T_1
    =
    \Theta\!\left(
        \frac{1}{\Delta}
        \log\frac{1}{\epsilon_1}
    \right).
\end{equation}
Therefore, the largest evolution time represented in the history state is
\begin{equation}
    T
    =
    T_1+T_2
    =
    \Theta\!\left(
        \frac{1}{\Delta}
        \log\frac{1}{\epsilon_1}
        +
        \frac{1}{\epsilon}
        \log\frac{1}{\delta}
    \right).
\end{equation}

As shown in Appendix~\ref{app:nonuniform-history-state}, the nonuniform history state in Eq.~\eqref{eq:spectral-history-state} can be prepared using
\begin{equation}
    \mytO{
        \alpha T
        \mylogpow{2}{
            \frac{1}{\epsilon_{\rm hist}}
        }
    }
\end{equation}
controlled queries to $U_K$, $U_0$, and their inverses. The construction is compatible with any suitable quantum linear system algorithm (QLSA), such as those of Refs.~\cite{costa2022optimal,jennings2023efficient,dalzell2024shortcut}.

Substituting the expression for $T$, the total query complexity is therefore
\begin{equation}
    \mytO{
        \alpha
        \left(
            \frac{1}{\Delta}
            \log\frac{1}{\epsilon_1}
            +
            \frac{1}{\epsilon}
            \log\frac{1}{\delta}
        \right)
        \mylogpow{2}{
            \frac{1}{\epsilon_{\rm hist}}
        }
    }.
\end{equation}

Suppose that $U_K$ and $U_0$ can each be implemented using
$\myO{\mypoly{\log M}}$ primitive quantum gates. Then the corresponding gate complexity is
\begin{equation}
    \mytO{
        \alpha
        \left(
            \frac{1}{\Delta}
            \log\frac{1}{\epsilon_1}
            +
            \frac{1}{\epsilon}
            \log\frac{1}{\delta}
        \right)
        \mylogpow{2}{
            \frac{1}{\epsilon_{\rm hist}}
        }
        \cdot
        \mypoly{\log M}
    }.
\end{equation}

The Kaiser window state can be prepared in
\begin{equation}
    \myO{J}
    =
    \myO{
        \frac{
            \alpha\log(1/\delta)
        }{
            \epsilon
        }
    }
\end{equation}
time, which does not exceed the leading scaling of the history-state preparation procedure. The inverse quantum Fourier transform requires only
$\myO{\mypoly{\log J}}$ primitive gates.

To ensure that the ideal and actual distributions are $O(\zeta)$-close in total variation distance, it suffices to choose
\begin{equation}
    \epsilon_1
    =
    \myTheta{
        \sqrt{\eta}\,\zeta
    },
    \qquad
    \epsilon_{\rm hist}
    =
    \myTheta{\zeta}.
\end{equation}
Consequently, when $\eta=\Theta(1)$, the query complexity becomes
\begin{equation}
    \mytO{
        \alpha
        \left(
            \frac{1}{\Delta}
            \log\frac{1}{\zeta}
            +
            \frac{1}{\epsilon}
            \log\frac{1}{\delta}
        \right)
        \mylogpow{2}{
            \frac{1}{\zeta}
        }
    },
\end{equation}
which is the complexity stated in Result~\ref{res:spectral}.

\subsection{Nonuniform history-state preparation}
\label{app:nonuniform-history-state}

We now describe a procedure for preparing a nonuniform history state of a
linear ODE. Consider
\begin{equation}
    \dot{x}(t)=Ax(t),
    \qquad
    x(t)=e^{At}x(0),
\end{equation}
where $A$ is normal and all its eigenvalues have non-positive real parts.
Suppose that $\|x(0)\|=1$, and that we have access to a block-encoding $U_A$
of $A$ with scale factor $\alpha_A\ge \|A\|$ and a unitary $U_0$ preparing
$\ket{x(0)}$.

Let 
\begin{align}
t_j=T_1+j\Delta t, \qquad j=0,1,\dots,J-1,    
\end{align}
and suppose that $\{w_j\}_{j=0}^{J-1}$ satisfy
$\sum_{j=0}^{J-1}|w_j|^2=1$,
with the state $\sum_{j=0}^{J-1}w_j\ket{j}$
being efficiently preparable. Our goal is to prepare the normalized version
of the nonuniform history state
\begin{equation}
    \ket{H_w}
    \defeq
    \sum_{j=0}^{J-1}
    w_j\ket{j}\ket{x(t_j)}.
    \label{eq:general-nonuniform-history-state}
\end{equation}
The coefficients $w_j$ may be chosen arbitrarily, subject to efficient state
preparation. In the Spectral-QKA algorithm, we take $w_j=\beta_j$, where
$\beta_j$ are the Kaiser-window coefficients.

\vspace{1em}
\paragraph{Linear-ODE solver via linear systems.}

We first recall the linear-system embedding of
Ref.~\cite{berry2024quantum}. For integers $m,p,l>0$ and time step $h>0$,
define
\begin{align}
C_{m,p,l}(Ah)
=
\sum_{s=0}^{m+p-1}
\ket{s}\bra{s}\otimes I
-
\sum_{s=0}^{m-1}
\ket{s+1}\bra{s}\otimes T_l(Ah)
-
\sum_{s=m}^{m+p-2}
\ket{s+1}\bra{s}\otimes I,
\end{align}
where
\begin{equation}
    T_l(Ah)
    =
    \sum_{r=0}^{l}\frac{(Ah)^r}{r!}
    \approx
    e^{Ah}.
\end{equation}
For the right-hand side
\begin{equation}
    \ket{b}
    =
    \ket{0}\ket{x(0)},
\end{equation}
the solution of
\begin{equation}
    C_{m,p,l}(Ah)\ket{y}
    =
    \ket{b}
\end{equation}
is
\begin{align}
    \ket{y}
    =
    \sum_{s=0}^{m-1}
    \ket{s}\,
    T_l(Ah)^s\ket{x(0)}
    +
    \sum_{s=m}^{m+p-1}
    \ket{s}\,
    T_l(Ah)^m\ket{x(0)}.
    \label{eq:linear-system-history-state}
\end{align}
Thus, the first $m$ blocks encode the time evolution, while the remaining
$p$ blocks contain repeated copies of the approximate final-time state.

We will choose the time step $h$ below such that
\begin{equation}
    h\le \frac{1}{\alpha_A}.
\end{equation}
Since $\alpha_A\ge \|A\|$, this ensures that
$h\|A\|=O(1)$. We choose $\epsilon_{\rm disc}\in(0,1)$ and the
truncation order $l$ such that the one-step approximation error satisfies
\begin{equation}
    \left\|
        T_l(Ah)-e^{Ah}
    \right\|
    \le
        \frac{\epsilon_{\rm disc}}{2m}.
\end{equation}
Since $\|e^{Ah}\|\le 1$, this implies, uniformly for
$s=0,1,\dots,m$,
\begin{equation}
    \left\|
        T_l(Ah)^s-e^{Ash}
    \right\|
    \le
    \epsilon_{\rm disc},
    \qquad
    \|T_l(Ah)^s\|=O(1).
    \label{eq:history-discretization-error}
\end{equation}
As in Ref.~\cite{berry2024quantum}, it suffices to take
\begin{equation}
    l
    =
    \mytO{
        \mylog{
            \frac{m}{\epsilon_{\rm disc}}
        }
    }.
    \label{eq:history-taylor-order}
\end{equation}
Consequently,
\begin{equation}
    \|C_{m,p,l}(Ah)\|
    =
    O(1),
    \qquad
    \|C_{m,p,l}(Ah)^{-1}\|
    =
    O(m+p).
    \label{eq:C-condition-number}
\end{equation}
A block-encoding of $C_{m,p,l}(Ah)$ with scale factor $\Theta(1)$ can be
implemented using $O(l)$ controlled queries to $U_A$ and its inverse.

\vspace{1em}
\paragraph{Uniform-size construction for multiple target times.}

We next construct linear systems of the same dimension for all target times.
Let
\begin{equation}
    T_2=(J-1)\Delta t,
    \qquad
    T=T_1+T_2,
\end{equation}
and choose
\begin{equation}
    m_\star=\lceil \alpha_A T\rceil,
    \qquad
    h=\frac{T}{m_\star}\le \frac{1}{\alpha_A}.
\end{equation}
For simplicity, assume that
\begin{equation}
    T_1=ah,
    \qquad
    \Delta t=ch
\end{equation}
for integers $a,c>0$. Rounding the target times to the nearest grid points
does not affect the asymptotic complexity. These relations imply
\begin{equation}
    m_\star=a+(J-1)c.
\end{equation}

For each $j=0,1,\dots,J-1$, define
\begin{equation}
    m_j=a+jc,
\end{equation}
so that
\begin{equation}
    m_jh=t_j.
\end{equation}
Since $m_j\le m_\star$ for every $j$, we choose a common truncation
order $l$ according to Eq.~\eqref{eq:history-taylor-order} with
$m=m_\star$. Then Eq.~\eqref{eq:history-discretization-error} holds
for every $m_j$.

We then choose
\begin{equation}
    p_j
    =
    2m_\star-m_j
    =
    a+(2J-2-j)c.
    \label{eq:def-mj-pj-history}
\end{equation}
Hence
\begin{equation}
    m_j+p_j=2m_\star
\end{equation}
is independent of $j$.

The solution of 
\begin{equation}
    C_{m_j,p_j,l}(Ah)\ket{y^{(j)}}
    =
    \ket{b}
\end{equation}
is
\begin{align}
    \ket{y^{(j)}}
    =
    \sum_{s=0}^{m_j-1}
    \ket{s}\,
    T_l(Ah)^s\ket{x(0)}
    +
    \sum_{s=m_j}^{2m_\star-1}
    \ket{s}\,
    T_l(Ah)^{m_j}\ket{x(0)}.
    \label{eq:yj-uniform-history}
\end{align}

The key property of this construction is that the same set of latter blocks,
namely
\begin{equation}
    s=m_\star,m_\star+1,\dots,2m_\star-1,
\end{equation}
lies in the padding region for every $j$. Thus, these $m_\star$ blocks
contain the same approximate final-time state
$T_l(Ah)^{m_j}\ket{x(0)}$ within each branch.

\vspace{1em}
\paragraph{Multiplexed linear system.}

We now combine all target times into a single linear system. Define
\begin{equation}
    \mathcal{C}
    \defeq
    \sum_{j=0}^{J-1}
    \ket{j}\bra{j}
    \otimes
    C_{m_j,p_j,l}(Ah),
    \label{eq:multiplexed-C}
\end{equation}
and prepare the right-hand side
\begin{equation}
    \ket{B}
    \defeq
    \sum_{j=0}^{J-1}
    w_j\ket{j}\ket{b}.
    \label{eq:multiplexed-b}
\end{equation}
Since $\sum_j|w_j|^2=1$ and $\|b\|=1$, the state $\ket{B}$ is normalized.

By linearity,
\begin{align}
    \ket{Y}
    \defeq
    \mathcal{C}^{-1}\ket{B}
    =
    \sum_{j=0}^{J-1}
    w_j\ket{j}\ket{y^{(j)}}.
    \label{eq:multiplexed-solution}
\end{align}
Thus, a single QLSA applied to
$\mathcal{C}\ket{Y}=\ket{B}$ prepares all target-time histories coherently,
with their desired relative weights already incorporated in the right-hand
side.

Since $\mathcal{C}$ is block diagonal,
\begin{equation}
    \|\mathcal{C}\|
    =
    \max_j
    \|C_{m_j,p_j,l}(Ah)\|
    =
    O(1),
\end{equation}
while
\begin{equation}
    \|\mathcal{C}^{-1}\|
    =
    \max_j
    \|C_{m_j,p_j,l}(Ah)^{-1}\|
    =
    O(m_\star).
\end{equation}
Hence the condition number of the multiplexed linear system is
\begin{equation}
    \kappa_{\mathcal C}
    =
    O(m_\star)
    =
    O(\alpha_A T).
    \label{eq:multiplexed-condition-number}
\end{equation}

The block-encoding of $\mathcal{C}$ can be implemented with essentially the
same cost as that of a single $C_{m_j,p_j,l}(Ah)$. Indeed, for given values
of $j$ and $s$, one can reversibly determine whether $s<m_j$ and thereby
select either the evolution block $-T_l(Ah)$ or the padding block $-I$.
Consequently, a block-encoding of $\mathcal{C}$ with scale factor
$\Theta(1)$ can be implemented using $O(1)$ controlled applications of a
block-encoding of $T_l(Ah)$, and hence using $O(l)$ controlled queries to
$U_A$ and its inverse.

\vspace{1em}
\paragraph{Extraction of the nonuniform history state.}

Let
\begin{equation}
    P_{\rm pad}
    \defeq
    \sum_{s=m_\star}^{2m_\star-1}
    \ket{s}\bra{s}
\end{equation}
be the projector onto the common padding region. Also define the discretized
nonuniform history state
\begin{equation}
    \ket{H_w^{(l)}}
    \defeq
    \sum_{j=0}^{J-1}
    w_j\ket{j}
    T_l(Ah)^{m_j}\ket{x(0)}.
    \label{eq:discrete-nonuniform-history}
\end{equation}
Applying $P_{\rm pad}$ to the block register of
Eq.~\eqref{eq:multiplexed-solution} gives
\begin{align}
    (I\otimes P_{\rm pad}\otimes I)\ket{Y}
    =
    \sum_{j=0}^{J-1}
    w_j\ket{j}
    \left(
        \sum_{s=m_\star}^{2m_\star-1}\ket{s}
    \right)
    T_l(Ah)^{m_j}\ket{x(0)}.
    \label{eq:projected-multiplexed-solution}
\end{align}
Define
\begin{equation}
    \ket{u_{\rm pad}}
    \defeq
    \frac{1}{\sqrt{m_\star}}
    \sum_{s=m_\star}^{2m_\star-1}\ket{s}.
\end{equation}
Then
\begin{equation}
    (I\otimes P_{\rm pad}\otimes I)\ket{Y}
    =
    \sqrt{m_\star}\,
    \ket{u_{\rm pad}}
    \sum_{j=0}^{J-1}
    w_j\ket{j}
    T_l(Ah)^{m_j}\ket{x(0)},
\end{equation}
where on the right-hand side we have swapped the first two registers for
notational convenience. The padding register therefore factors from the remaining registers and is
in the fixed state $\ket{u_{\rm pad}}$, independent of the branch index $j$.
For the exact solution, unpreparing $\ket{u_{\rm pad}}$ and postselecting the
padding register onto $\ket{0}$ succeeds with certainty and leaves the
normalized version of $\ket{H_w^{(l)}}$ in the remaining registers.

By Eq.~\eqref{eq:history-discretization-error},
\begin{align}
    \|
        \ket{H_w^{(l)}}-\ket{H_w}
    \|^2
    =
    \sum_{j=0}^{J-1}
    |w_j|^2
    \left\|
        T_l(Ah)^{m_j}\ket{x(0)}
        -
        \ket{x(t_j)}
    \right\|^2
    \le
    \epsilon_{\rm disc}^2.
    \label{eq:nonuniform-history-discretization}
\end{align}

\vspace{1em}
\paragraph{Success probability and error in the Koopman setting.}

We now specialize to the Koopman evolution considered in this section,
namely $A=K$ and $x(t)=g(t)$. Since $K$ is normal and all its eigenvalues
have non-positive real parts,
\begin{equation}
    \|g(t)\|
    \le
    \|g(0)\|
    =
    1,
    \qquad
    t\ge0.
\end{equation}
Moreover,
\begin{align}
    \|g(t)\|^2
    =
    \|\Pi_S g(t)\|^2
    +
    \|\Pi_F g(t)\|^2
    \ge
    \|\Pi_S g(t)\|^2
    =
    \|\Pi_S g(0)\|^2
    =
    \|a_S\|^2
    \ge
    \eta.
    \label{eq:koopman-norm-lower-bound}
\end{align}
Thus,
\begin{equation}
    \sqrt{\eta}
    \le
    \|g(t)\|
    \le
    1,
    \qquad
    t\ge0.
\end{equation}

Meanwhile, since $\sum_{j=0}^{J-1}|w_j|^2=1$,
\begin{equation}
    \|\ket{H_w}\|^2
    =
    \sum_{j=0}^{J-1}
    |w_j|^2\|g(t_j)\|^2
    \ge
    \eta.
\end{equation}
Combining this with
Eq.~\eqref{eq:nonuniform-history-discretization}, we obtain
\begin{equation}
    \|\ket{H_w^{(l)}}\|
    \ge
    \sqrt{\eta}-\epsilon_{\rm disc}.
\end{equation}
Hence, for sufficiently small
$\epsilon_{\rm disc}\le c\sqrt{\eta}$ with some constant $c<1$,
\begin{equation}
    \|\ket{H_w^{(l)}}\|^2
    =
    \Omega(\eta).
\end{equation}

On the other hand, every branch $\ket{y^{(j)}}$ contains $2m_\star$ blocks,
each of norm $O(1)$, and therefore
\begin{equation}
    \|\ket{Y}\|^2
    =
    \sum_{j=0}^{J-1}
    |w_j|^2
    \|\ket{y^{(j)}}\|^2
    =
    O(m_\star).
\end{equation}
It follows that the probability of projecting onto the common padding region
is
\begin{align}
    p_{\rm pad}
    =
    \frac{
        \|
            (I\otimes P_{\rm pad}\otimes I)\ket{Y}
        \|^2
    }{
        \|\ket{Y}\|^2
    }
    =
    \frac{
        m_\star
        \|\ket{H_w^{(l)}}\|^2
    }{
        \|\ket{Y}\|^2
    }
    =
    \myOmega{\eta}.
    \label{eq:history-postselection-probability}
\end{align}
In particular, under the assumption $\eta=\myOmega{1}$ used in
Result~\ref{res:spectral}, the post-selection succeeds with constant
probability.

Suppose that the QLSA prepares a state
$\ket{\widehat{Y}}$ satisfying
\begin{equation}
    \|
        \ket{\widehat{Y}}
        -
        \ket{\bar Y}
    \|
    \le
    \epsilon_{\rm qlsa},
\end{equation}
where $\epsilon_{\rm qlsa}=O(\sqrt{\eta})$ is sufficiently small.
Conditioning on the common padding region changes the state error by at most
a factor $O(1/\sqrt{\eta})$. The resulting normalized state is therefore
$O(\epsilon_{\rm qlsa}/\sqrt{\eta})$-close to
$\ket{u_{\rm pad}}\ket{\bar H_w^{(l)}}$. After unpreparing
$\ket{u_{\rm pad}}$, postselection of the padding register onto $\ket{0}$
succeeds with probability
\begin{equation}
    1-
    O\!\left(
        \frac{\epsilon_{\rm qlsa}^2}{\eta}
    \right),
\end{equation}
and leaves the remaining registers
$O(\epsilon_{\rm qlsa}/\sqrt{\eta})$-close to
$\ket{\bar H_w^{(l)}}$.

Combining this with
Eq.~\eqref{eq:nonuniform-history-discretization}, we obtain a final state
$\ket{\widehat H_w}$ satisfying
\begin{equation}
    \|
        \ket{\widehat H_w}
        -
        \ket{\bar H_w}
    \|
    =
    O\!\left(
        \frac{
            \epsilon_{\rm disc}
            +
            \epsilon_{\rm qlsa}
        }{
            \sqrt{\eta}
        }
    \right).
    \label{eq:nonuniform-history-total-error}
\end{equation}
It therefore suffices to choose
\begin{equation}
    \epsilon_{\rm disc}
    =
    \myTheta{
        \sqrt{\eta}\,
        \epsilon_{\rm hist}
    },
    \qquad
    \epsilon_{\rm qlsa}
    =
    \myTheta{
        \sqrt{\eta}\,
        \epsilon_{\rm hist}
    },
\end{equation}
to obtain
\begin{equation}
    \|
        \ket{\widehat H_w}
        -
        \ket{\bar H_w}
    \|
    =
    O(\epsilon_{\rm hist}).
\end{equation}

\vspace{1em}
\paragraph{Query complexity.}

The construction above can be combined with any QLSA 
that prepares an approximation to the normalized solution of
$\mathcal{C}\ket{Y}=\ket{B}$. For example, using a QLSA with query complexity
\begin{equation}
    \mytO{
        \kappa_{\mathcal C}
        \mylog{
            \frac{1}{\epsilon_{\rm qlsa}}
        }
    }
\end{equation}
in terms of queries to a block-encoding of $\mathcal C$ and to the
right-hand-side preparation unitary, together with
Eqs.~\eqref{eq:history-taylor-order} and
\eqref{eq:multiplexed-condition-number}, gives
\begin{equation}
    \mytO{
        \alpha_A T
        \mylogpow{2}{
            \frac{1}{\epsilon_{\rm hist}}
        }
    }
    \label{eq:nonuniform-history-query-complexity}
\end{equation}
controlled queries to $U_A$, $U_0$, and their inverses when
$\eta=\Theta(1)$. Here logarithmic dependence on $\alpha_A T$ is absorbed
into the $\widetilde O$ notation.

Therefore, in the Koopman setting $A=K$, the state
\begin{equation}
    \ket{H}
    =
    \sum_{j=0}^{J-1}
    \beta_j
    \ket{j}
    \ket{g(T_1+j\Delta t)}
\end{equation}
can be prepared, after normalization, to accuracy
$O(\epsilon_{\rm hist})$ using
\begin{equation}
    \mytO{
        \alpha T
        \mylogpow{2}{
            \frac{1}{\epsilon_{\rm hist}}
        }
    }
\end{equation}
controlled queries to $U_K$, $U_0$, and their inverses, as used in the Spectral-QKA algorithm.

\end{document}